\documentclass[a4paper,german,twocolumn]{article}
\usepackage{a4}
\usepackage{vmargin}
\usepackage{latexsym}

\usepackage{graphicx}
\usepackage{float}
\usepackage[vflt]{floatflt}

\usepackage{charter}

\newcommand{\vspm}{\vspace{-1cm}}

\newcommand{\mb}[1]{\mathbf{#1}}
\newcommand{\bm}[1]{\mbox{\boldmath$#1$\unboldmath}}
\newcommand{\tm}[1]{\textrm{#1}}

\newcommand{\ind}[1]{{\textrm{\tiny{#1}}}}
\newcommand{\eww}[1]{\left\la{#1}\right\ra}

\newcommand{\gradient}{\bm{\partial}}

\newcommand{\la}{\langle}
\newcommand{\ra}{\rangle}
\renewcommand{\phi}{\varphi}
\newcommand{\phithreed}{\phi_{3d}}
\newcommand{\phitwod}{\phi_{2d}}

\newcommand{\sigPoly}{\sigma_{\rm P}}
\newcommand{\bcfig}{\begin{figure}[H]\begin{center}}
\newcommand{\ecfig}{\end{center}\end{figure}}

\newcommand{\var}{\tm{Var}}
\newcommand{\Var}[1]{{\tm{Var}}\left[#1\right]}

\newcommand{\ta}{\tau_\alpha}
\newcommand{\Mta}{$\tau_\alpha$\ }

\newcommand{\sqrr}{\langle r^2(t)\rangle}

\newcommand{\Nco}[1]{N_{\ind{#1}}^{\ind{coop}}}
\newcommand{\Ncodv}{\Nco{$\delta v$}}
\newcommand{\Ncon}{\Nco{$\mb{n}$}}
\newcommand{\Ncov}{\Nco{$\mb{v}$}}
\newcommand{\NcoX}{\Nco{$X$}}

\newcommand{\Ncovi}{\Nco{$\mb{v},\!\infty$}}
\newcommand{\Ncovr}{\Nco{$\mb{v},$ren}}

\newcommand{\Ncodvmax}{\Nco{$\delta v,$max}}
\newcommand{\Nconmax}{\Nco{$\mb{n},$max}}

\newcommand{\Ncoop}{\NcoX}
\newcommand{\MNcoop}{$\Ncoop$\ }

\newcommand{\SX}{S_{X}(R,t)}
\newcommand{\Sdv}{S_{\delta v}(R,t)}
\newcommand{\Svv}{S_{\mb{v}}(R,t)}
\newcommand{\Sn}{S_{\mb{n}}(R,t)}
\newcommand{\xiX}{\xi_X}
\newcommand{\xidv}{\xi_{\delta v}}
\newcommand{\xiv}{\xi_{\mb{v}}}
\newcommand{\xin}{\xi_{\mb{n}}}
\newcommand{\xiXi}{\xi_X(\!\infty\!)}
\newcommand{\xidvi}{\xi_{\delta v}(\!\infty\!)}
\newcommand{\xivi}{\xi_{\mb{v}}(\!\infty\!)}
\newcommand{\xini}{\xi_{\mb{n}}}
\newcommand{\Sperp}{S_{\mb{n}}^{\perp}(R,t)}
\newcommand{\Spara}{S_{\mb{n}}^{||}(R,t)}
\newcommand{\xiperp}{\xi_{\mb{n}}^{\perp}}
\newcommand{\xipara}{\xi_{\mb{n}}^{||}}
\newcommand{\Rcurl}{R_\ind{curl}}
\newcommand{\Rcurli}{R_\ind{curl}(\infty)}

\newcommand{\vi}{\mb{v}_i}
\newcommand{\vj}{\mb{v}_j}

\newcommand{\ui}{\mb{u}_i}
\newcommand{\wi}{\mb{w}_i}
\newcommand{\Dv}[2]{\mb{\Delta}_{#1}^{#2}}
\newcommand{\vcm}{\mb{v}_\ind{cm}}

\newcommand*{\figany}[4]{
\begin{figure}[#1]
\begin{center}
     #2
   \caption{\label{#4}#3}
\end{center}
\end{figure}
}

\newcommand{\NEQU}[2]
{
\begin{equation}
     #1
\label{#2}
\end{equation}
}

\newcommand{\EQU}[2]
{
$$      #1	$$
}

\begin{document}

\newcommand{\FIG}[1]{Fig. \ref{#1}}
\newcommand{\FIGU}[1]{Figure \ref{#1}}
\newcommand{\REQU}[1]{Equ.(\ref{#1})}

\onecolumn

\title{Cooperativity and Spatial Correlations near the Glass Transition:
Computer Simulation Results for Hard Spheres and Discs}

\author{B. Doliwa\thanks{email: doliwa@mpip-mainz.mpg.de}\ \ \ and A. Heuer}
\date{Max-Planck-Institut f\"ur Polymerforschung, Postfach 3148\\ D-55021 Mainz, Germany}
\maketitle
\begin{abstract}
We examine the dynamics of hard spheres
and discs at high packing fractions
in two and three dimensions,
modeling the simplest systems
exhibiting a glass transition.
As it is well known, cooperativity and dynamic heterogeneity
arise as central features when approaching the glass transition
from the liquid phase, so an understanding of their underlying
physics is of great interest.
Cooperativity implies a reduction of 
the effective degrees of freedom, and we demonstrate a simple way
of quantification in terms of the strength and the length scale of dynamic 
correlations among different particles. These correlations are obtained for different
dynamical quantities $X_i(t)$ that are
constructed from single-particle displacements during some observation time $t$.
Of particular interest is the dependence on $t$.
Interestingly, for appropriately chosen
$X_i(t)$ we obtain finite cooperativity in the limit $t \rightarrow \infty$. 
\end{abstract}
\twocolumn
The remarkable features of glass-forming liquids, so is
agreed to presently, can widely be attributed to collective phenomena.
We know that they become more and more important if we approach
the glass-transition point. It is a great
challenge to understand collective phenomena
because we expect them to be a
kind of universal origin of glassy behavior.

The goal of this paper is to introduce a way of quantifying
the degree of cooperativity
by exploiting the correlations of two-time, single-particle quantities.
This will be done for two- and three-dimensional systems
thus revealing their very similar behavior.
Naturally, these correlations have a spatial aspect,
which shall be examined in detail.
In principle, the idea is not new, because a considerable amount
of work has recently been done on this subject
\cite{Poole:1998,Bennemann:1999,KobGlotzerPRE:1999,Yamamoto:1998}.
Even an experimental determination of dynamical length scales
has become possible through multi-dimensional NMR
\cite{Tracht:1998,Tracht:1999}.
Particularly interesting, the theory of spin glasses
makes predictions about
the behavior of dynamic susceptibilities and - connected to it -
dynamical length scales, when approaching the glass transition
\cite{Parisi:1999}.
Summing spatial correlations, the susceptibility shows a divergence
for the analyzed spin model near the
mode-coupling critical temperature $T_c^+$.
There is some evidence, that this divergence is present
in structural glasses, too
\cite{Donati:1999,Novikov:1999}.

Up to now, the discussion of dynamical length scales has mainly been focussed
on the mobility of particles, but not on their direction of motion.
Various versions of spatial correlators are in use, e.g.
$\la\mu(0,t)\mu(R,t)\ra$, where $\mu_i(t)$ represents the length of the
total displacement during $[0,t]$, i.e. the mobility, see \cite{Donati:1999}.
As another example, one attributes to $\mu_i(t)$ the value of one, if
particle $i$ is slow, and zero else \cite{Novikov:1999}, see section 2.
We will demonstrate, however, that the directional aspect of motion
is crucial for interparticle correlations, resulting, e.g., in a much stronger
density dependence of dynamical length scales.

From correlators of the above type, we can obtain the spatial extent
as well as the overall, or mean, degree of cooperativity in the system's
motion.
In literature, one uses the detour via Fourier space and the
fitting of Orstein-Zernike functions, to determine
dynamical length scales. Throughout the present work, however, we will
stay in real space, which will ease the interpretation of our data.

We will present a treatment of the overall cooperativity,
which, in contrast to existing work, makes possible a pictorial understanding
of our results in terms of a reduction of degrees of freedom.
An important point is the time scale $t$ defining the dynamical measurements.
We hope to demonstrate, first, that dynamical length scales strongly depend on $t$,
and, second, that $t=\ta$ is not a sensible choice. Furthermore we show that
the reduction of the degrees of freedom is directly related to the Haven ratio, well known to characterize
cooperativity effects for the ion dynamics in ion conductors
\cite{Maass:1995}.

The organization of the paper is as follows.
Section 1 gives the details of the performed simulations and
introduces the main dynamical features via common
single-particle quantities.
Section 2 formulates our approach to quantify the system's
overall cooperativity, which is well-known to be the
integral of spatial correlations.
The latter are treated in section 3, obtaining their strength and
length scale.
We conclude by a discussion of our results in section 4.

\section{Simulation details}

It is the advantage of a hard-sphere (HS) system for computer experiments,
that the pair potential
\EQU{V_{ij}(r_{ij})=\left\{\begin{array}{r@{\quad:\quad}l}
		\infty & r_{ij}<R_i+R_j \\ 0 & \tm{otherwise}\end{array}\right.}{}
forbids certain regions of configurational space, so that Monte Carlo steps
are simply denied when particle overlaps occur.
These computer-friendly 'yes/no' decisions make a Monte Carlo implementation
of HS dynamics very efficient.
The volume fraction
\EQU{\phithreed\equiv\frac1V\sum\frac43\pi R_i^3, \tm{ for $d\!=\!3$}}{}
or
\EQU{\phitwod\equiv\frac1V\sum\pi R_i^2, \tm{ for $d\!=\!2$}}{}
takes the role of temperature, which is not a relevant control parameter here.

An important input parameter is the distribution of particle
sizes $R_i$, i.e. the polydispersity.
It determines to a large extent, how amorphous the system is.
For example, a bimodal mixture of spheres can be used to prevent
crystallization. In this work, we use a continuous, gaussian
distribution of width $\sigPoly$ and mean radius $\la R_i\ra=1$,
the latter serving as the unit of length.
Particles of $|R_i-1|>3\sigPoly$ were not used, because they would
slow down the simulations very much.
Former experiments show, that for 3d systems, $\sigPoly$=$10\%$
is enough to obtain a stable amorphous state, i.e. lacking
long-range order \cite{Moriguchi:1993}.
In the two-dimensional case (discs), we will work with
$\sigPoly=25\%$.

Although a HS system seems rather artificial at first sight,
there is great interest in its properties, from both
the theoretical and the experimental side.
That, on the one hand, is due to the unbeatable simplicity,
and on the other hand
to the fact that HS are well represented by colloids
in real life. Microscopically, colloidal particles perform
free diffusion in their solvent, which is
one of the reasons why we have chosen a Monte Carlo
algorithm to generate the dynamics.
Rather than integrating Newton's $F=ma$, we propagate the system
according to the Langevin equation
\NEQU{\zeta\dot{\mathbf{r}}_i=-\gradient_{i}V(\mb{r}_1,...\mb{r}_N)
			+ \bm{\eta}_i,}{XLANGEVIN}
where white noises $\bm{\eta}_i(t)$ are directly coupled to
the particles' {\it positions}.
For very short waiting times $t$, there will hardly be any
collision, so the
potential term in \REQU{XLANGEVIN} can be neglected.
The result is a free diffusion for $t\!\to\!0$, i.e. $\sqrr\approx2dD_0t$,
where $d\!\in\!\{2,3\}$ denotes the number of dimensions.
All particles have equal masses, and their microscopic diffusion
constants $D_0$ will be the same.
Although this kind of dynamics is convenient for simulations,
the approach via Newton's equations (molecular dynamics) would
lead to quite similar results.
Naturally, the trivial short-time motion would be completely different
from the Monte Carlo case, but
the relevant information for longer times is expected to be
insensitive to the microscopic dynamics.
Recently, this has been demonstrated for a Lennard-Jones type system
\cite{Gleim:1998}.

In a Monte Carlo step, we randomly
choose a particle and try to displace
it a random amount $d\mb{x}$, whose distribution has the width
$\lambda$. Thus, $\lambda$ is the typical step length.
We must take it as small as possible, because only in the
limit $\lambda\to0$, we are sure to
integrate \REQU{XLANGEVIN} correctly.
On the other hand, a too small $\lambda$ will reduce our simulation
efficiency extremely, because upon halving $\lambda$, we need four
times as many steps to cover the same distance.
As a compromise, we try to achieve an acceptance rate of 50 percent,
i.e. half of the displacements should result in valid moves, that is,
producing no particle overlaps.
Dependent on the packing fraction, this yield values from $\lambda=0.02$
to $\lambda=0.05$, i.e. a few percent of the mean particle radius.
Comparisons to simulation runs with much smaller step sizes
showed this choice to be sensible because no deviations arose,
except for a trivial shift of time axis.

In the simulations analyzed in this paper, we used relatively
large systems in order to prevent major finite size effects.
To be more specific, we have
$N=$8756, 8960, 9201 and 9320 for the two-dimensional systems at
$\phitwod=$0.73, 0.75, 0.77 and 0.78, which
correspond to a box length of a hundred mean particle diameters.
In the three-dimensional case, we used $N=$15422 and 16307 particles
filling a volume $(50R_0)^3$, i.e.  25 mean particle diameters
in each direction. The corresponding volume fractions
are $\phithreed=$0.53 and 0.56.

To get a first impression of the system's dynamics,
it is most simple to calculate two-time, one-particle quantities.
They show the same strong dependence on packing fraction as macroscopic
transport quantities, like viscosity, when approaching the glass
transition.
\figany{!ht}{\includegraphics[width=7.2cm]{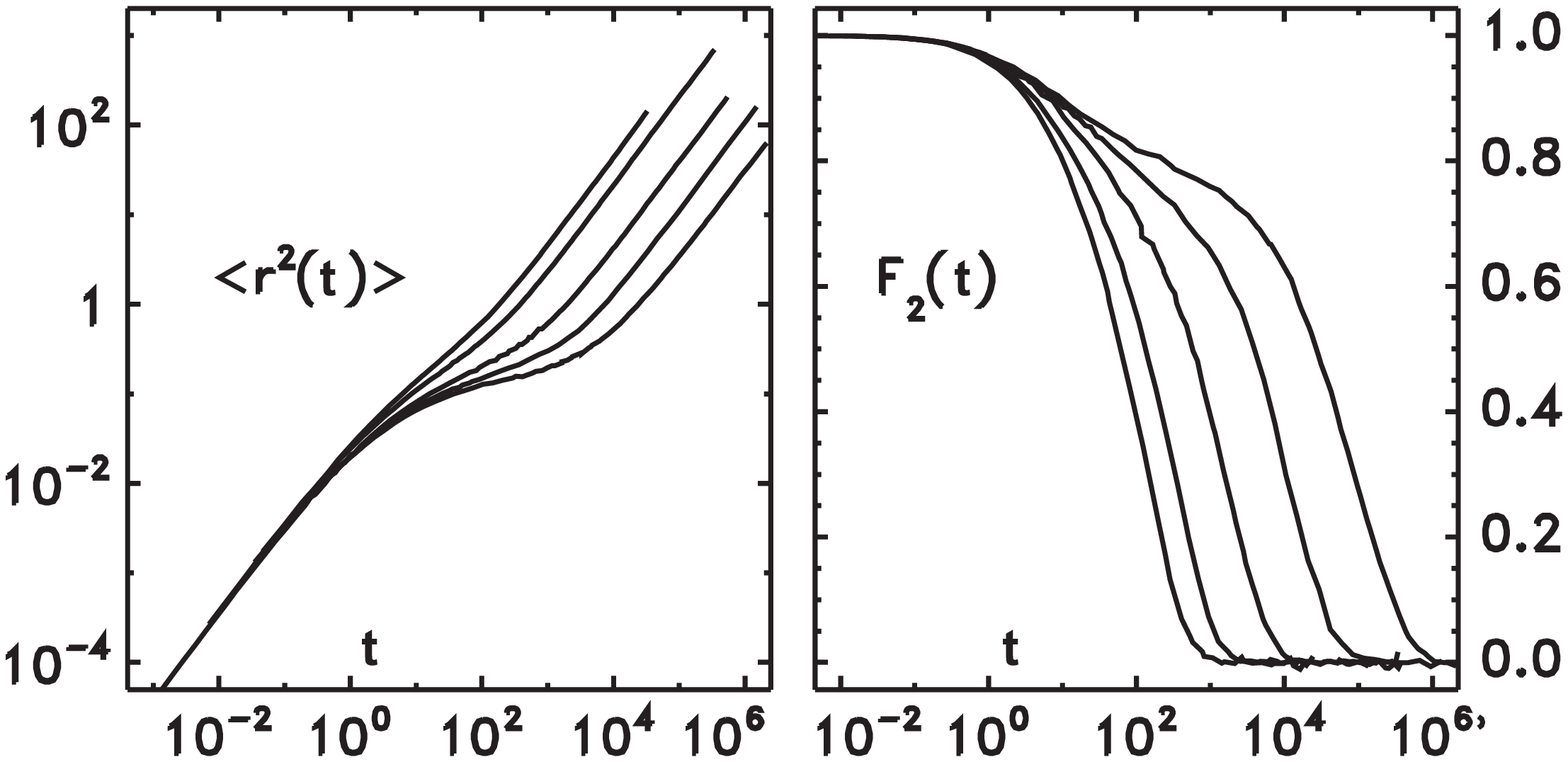}}{
One-particle, two-time quantities for the 3d packing fractions
$\phithreed=50\%, 53\%, 56\%, 57.3\%$ and $58\%$, from left to right.
The system sizes are $N\approx1000$.
(a) the mean squared displacement $\sqrr$, and
(b) the incoherent scattering function $F_2(\mb{k}_{\tiny\textnormal{max}},t)$.
}{SQRRS}
The mean squared displacement
\EQU{\sqrr\equiv\left\la\frac1N
\sum_{i}\left(\mb{r}_i(t)-\mb{r}_i(0)\right)^2\right\ra}{SQRR}
and the incoherent part of the
scattering function at a given wave vector $\mb{k}$,
\EQU{F_2(\mb{k},t)=\left\langle\frac1N\sum_{i}e^{i\mb{k}(\mb{r}_i(t)
	-\mb{r}_i(0))}\right\rangle
}{F2}
are the most common examples (see \FIG{SQRRS}).
The data shown in the figure above have been calculated in an earlier
work, using small 3d systems of $N\approx1000$ particles.
In the following, this data will not be used anymore.
Interestingly, the one-particle quantities in \FIG{SQRRS} differ little from
their counterparts in the large $N$ systems, i.e. for the analyzed packing
fractions $\phithreed=0.53$ and $0.56$.
The relaxation times $\ta$, for instance, agree within 20\% at $\phi=0.56$.
Many-particle correlations, in contrast, have turned out to be very sensitive to
system size, see below.
Simulation runs for $\phithreed>0.56$ with numbers of particles $N>10000$ are
not available
at the moment, but the densities $\phithreed=0.53$ and $0.56$ seem to
produce all the interesting features of a cold glass-forming liquid.

We want to emphasize at this point that the one-particle quantities for the
two-dimensional case look very similar to \FIG{SQRRS}, i.e. we find
anomalous diffusion, as expressed by the slope of $\sqrr$, and
a plateau in the scattering function $F_2$, when going to high densities
(not shown here).
We can extract from $\sqrr$ the ratio $\frac D{D_0}$ of
the long- and short-time diffusion constants.
It describes the slowing down of the particles' long-distance
motions upon increasing $\phi$.
\figany{!ht}{\includegraphics[width=7.2cm]{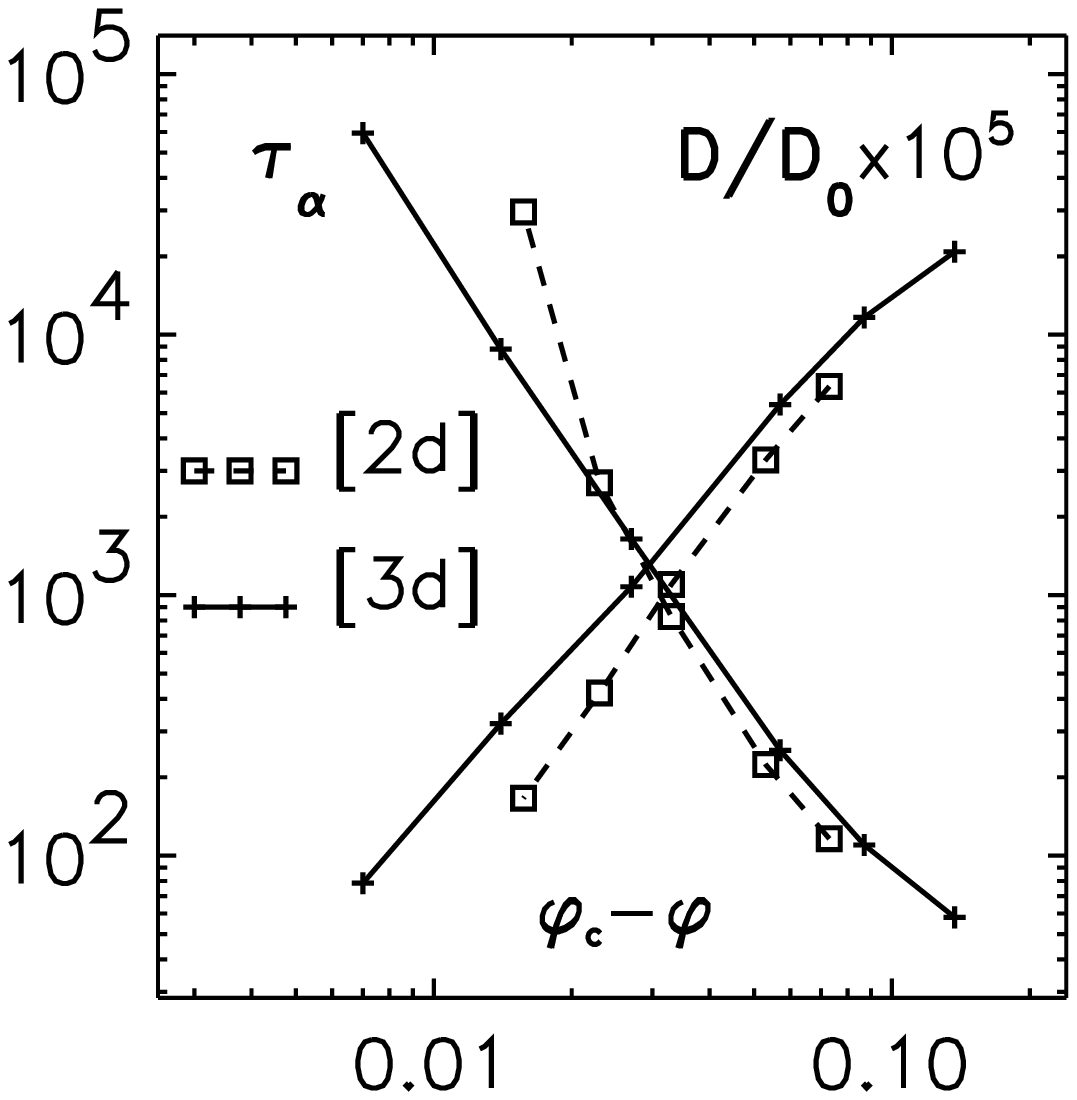}}
{Relaxation times $\ta(\phi)$ and diffusion constants $\frac D{D_0}(\phi)$
for 2d and 3d systems. The critical values $\phi_c$ are determined
by a MCT fit of the diffusion constant, i.e. $\frac{D}{D_0}\sim(\phi_c\!-\!\phi)^{-\gamma}$.
The results are $\phi_\ind{c,3d}=58.7\%$
and $\phi_\ind{c,2d}=80.3\%$.
}{TAUALPHAS}
The structural relaxation time $\ta$ is calculated according to
the condition
$F_2(\mb{k},\ta)=\frac1e$, where the
wave vector $\mb{k}=\mb{k}_{\tiny\textnormal{max}}$
corresponds to the next-neighbor distance.
The increasing $\ta$ and the decreasing $\frac D{D_0}$
indicate a great change of the dynamics when approaching
a critical value 
of $\phi_\ind{c,3d}=58.7\%$ in three
and $\phi_\ind{c,2d}=80.3\%$ in two dimensions, see \FIG{TAUALPHAS}.
We obtained $\phi_c$ by a fit of $\frac{D}{D_0}(\phi)$ to a power law
$\frac{D}{D_0}(\phi)\sim(\phi_c\!-\!\phi)^{-\gamma}$ as predicted by
mode coupling theory \cite{Gotze:1992}.
However, the exact value of $\phi_c$ should not be overinterpreted
because fits to a Vogel-Fulcher behavior
$\frac{D}{D_0}(\phi)\sim\exp(-\frac{\tm{C}}{\phi_c-\phi})$
work equally well, resulting in $\phi_\ind{VF,3d}=0.612$
and $\phi_\ind{VF,2d}=0.815$.

In the case of a HS system, the reason for the reduction of mobility
at high $\phi$ is quite easy to understand.
The particles are tightly surrounded by
the so called {\it cages} of next neighbors
which to a large degree restrict their motions.
On average, a particle feels a back-dragging force,
which prevents its cage from being destroyed,
\cite{doliwa:1998,doliwa:1999}.
If we define
$\mb{x}_i^{(m)}(\epsilon)\!\equiv
	\!\mb{r}_i(t\!=\!m\epsilon)\!-\!\mb{r}_i(t\!=\!m\epsilon\!-\!\epsilon)$
\newcommand{\xstep}[1]{\mb{x}^{(#1)}}
as the subsequent displacements of a tagged particle,
the back-dragging force results in a negative value of the correlation
$\la\xstep{1}\xstep{m}\ra$.
With the Green-Kubo relation
\EQU{D=D_0+\frac{1}{\epsilon d}\lim_{M\to\infty}
	\sum_{m=2}^{M}\la\xstep{1}\xstep{m}\ra,
}{GREENKUBO}
this immediately leads us to the conclusion that the so called
cage effect is responsible for the slowing down of the motion,
as expressed by $D<D_0$.

On a longer time scale, the particles finally succeed in leaving their
cages. Naturally, this can only occur if the neighbors rearrange in a
collective way.

\section{Cooperative Effects}

\FIGU{MOVES} suggests the fact that a
liquid near its glass transition possesses beautiful and
highly non-trivial dynamics,
see also \cite{Kolbe:1998,Perera:1999}.
\figany{!ht}{
\includegraphics[width=6cm]{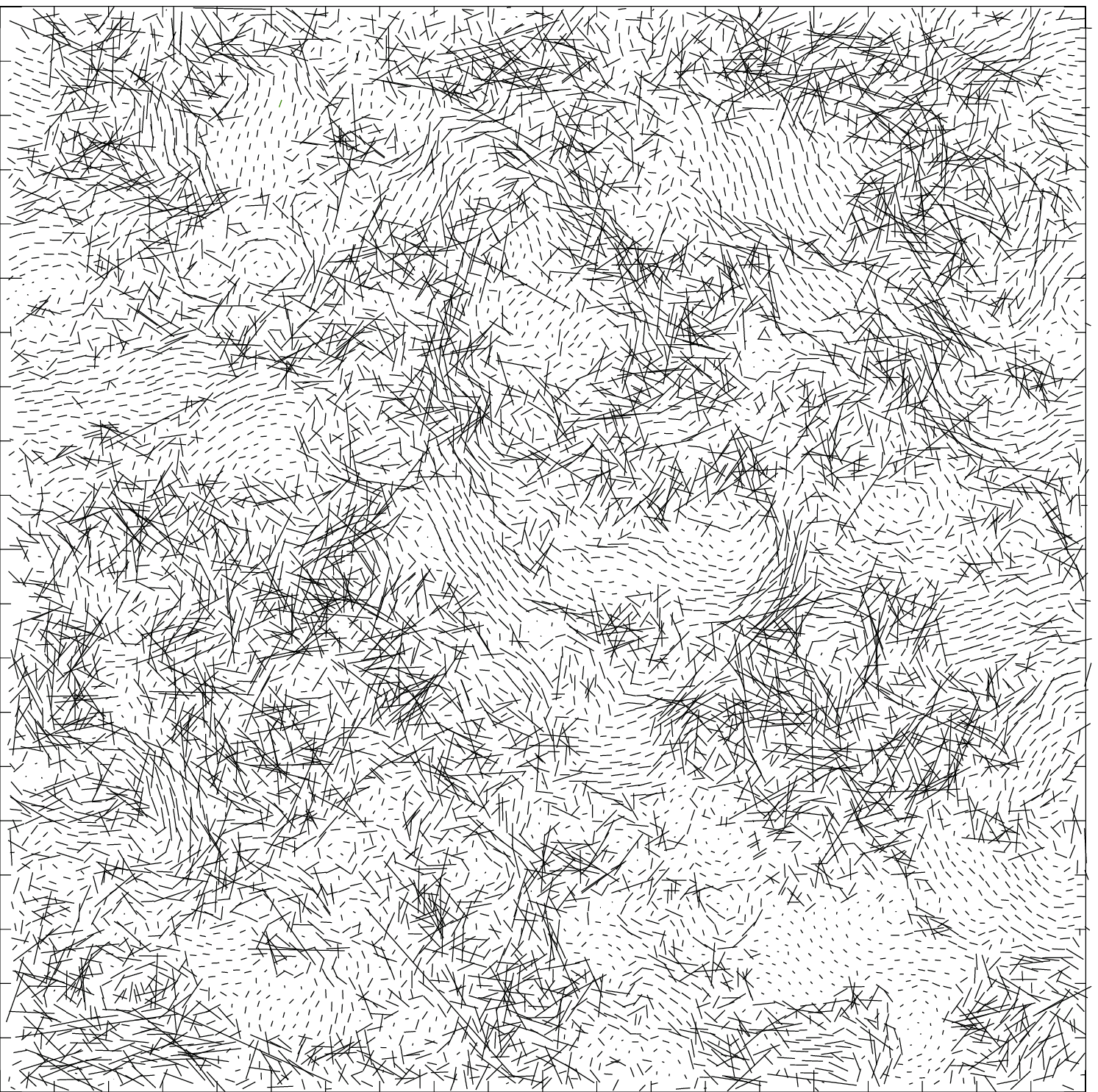}
}{Particle motions $\mb{r}_i(t+50\ta)-\mb{r}_i(t)$ during a time $50\ta$.
The density is $\phi=0.78$ in a 2d-system.}{MOVES}
We recognize regions of totally different behavior, some
of which show very crowded and uncooperative motions, while
others seem to act {\it{as one}} resulting in
collective flows.
The mobility obviously varies between different areas,
which is commonly referred to as {\it dynamic heterogeneity}.
How can we quantify the degree of cooperativity in our system?
It is possible to do this by comparing the fluctuation of
a one-particle dynamic quantity $X_i$ with
its many-particle equivalent $\sum_iX_i$.
For simplicity, let $\la X_i\ra=0$, which implies $\la\sum X_i\ra=0$.
We can think of $X_i$ to be the displacement
vector $X_i=\mb{v}_i\equiv\mb{r}_i(t+t')\!-\!\mb{r}_i(t)$ or its
relative length
$X_i=\delta v_i\equiv v_i-\la v_i\ra$,
where $v_i\equiv|\mb{v}_i|$.
The direction of motion $\mb{n}_i\equiv\mb{v}_iv_i^{-1}$ is a sensible
choice for $X_i$, too.
In the case where interparticle correlations are lacking,
the width of the distribution of $X$ will be
$$\Var{\sum X_i}=\sum\Var{X_i}.$$
Correlations, however, will increase $\Var{\sum X_i}$ while
anti-correlations will do the opposite.
From \FIG{MOVES}, it is reasonable to expect
{\it correlations}, and we define
\NEQU{\Ncoop\equiv\frac{\Var{\sum X_i}}{\sum\Var{X_i}}
	=1+\frac{\sum_{i\ne j}\la X_iX_j\ra}{\sum\la X_i^2\ra}.}{GLNCOOP}
If our expectation is right, then \MNcoop will be larger than one.

We now claim that \MNcoop measures the total reduction of degrees of
freedom caused by correlations.
In the simple case of uncorrelated motion ($\la X_iX_j\ra=0$),
we obtain $\Ncoop=1$, whereas the other extreme of totally correlated
motion ($X_i\equiv X_j$) results in $\Ncoop=N$.
If, more generally
we have $M$ identical
variables $X_i$ in each of $L$ independent
groups, $N=ML$, we obtain
\begin{eqnarray*}
\Ncoop &=& 1+\frac 1{\sum\la X_i^2\ra}\sum_{i=1}^N\sum_{j=1}^{M-1}\la X_{i}^2\ra=M.
\end{eqnarray*}
These examples show that $\Ncoop$ is indeed a reasonable quantitative
measure for the degree of cooperativity.
In real life, correlations will not be perfect, i.e. a hundred
percent, and their strengths and spatial extensions will vary throughout
the system. Hence, we should expect \MNcoop to be an average or effective
reduction factor for the degrees of freedom.

It is important to note that $\Ncoop=\Ncoop(t)$ because
the dynamic quantities $X_i=X_i(t)$ are dependent on the time scale
necessary for their definition.
The calculation of \MNcoop turns out to be quite unconvenient.
From one configuration, we only get one data point for the
term $\sum_iX_i$ in the numerator, i.e. \MNcoop is not self-averaging.
Thus, the simulation run has to be very long to aquire enough points
for the calculation of the variance.
A way around this obstacle is not to sum over the whole system,
but only over local subsystems of $n$ particles.
This should improve statistics. 
In practice, at a given $n$, we randomly choose a central particle
and add the closest
$(n-1)$ neighbors using them
as a subsystem. Repeating this procedure for a small number of
other central particles, we get some more subsystems of size $n$.

\figany{!ht}{
\begin{center} \hspace{-2cm}
\centering\includegraphics[height=7cm]{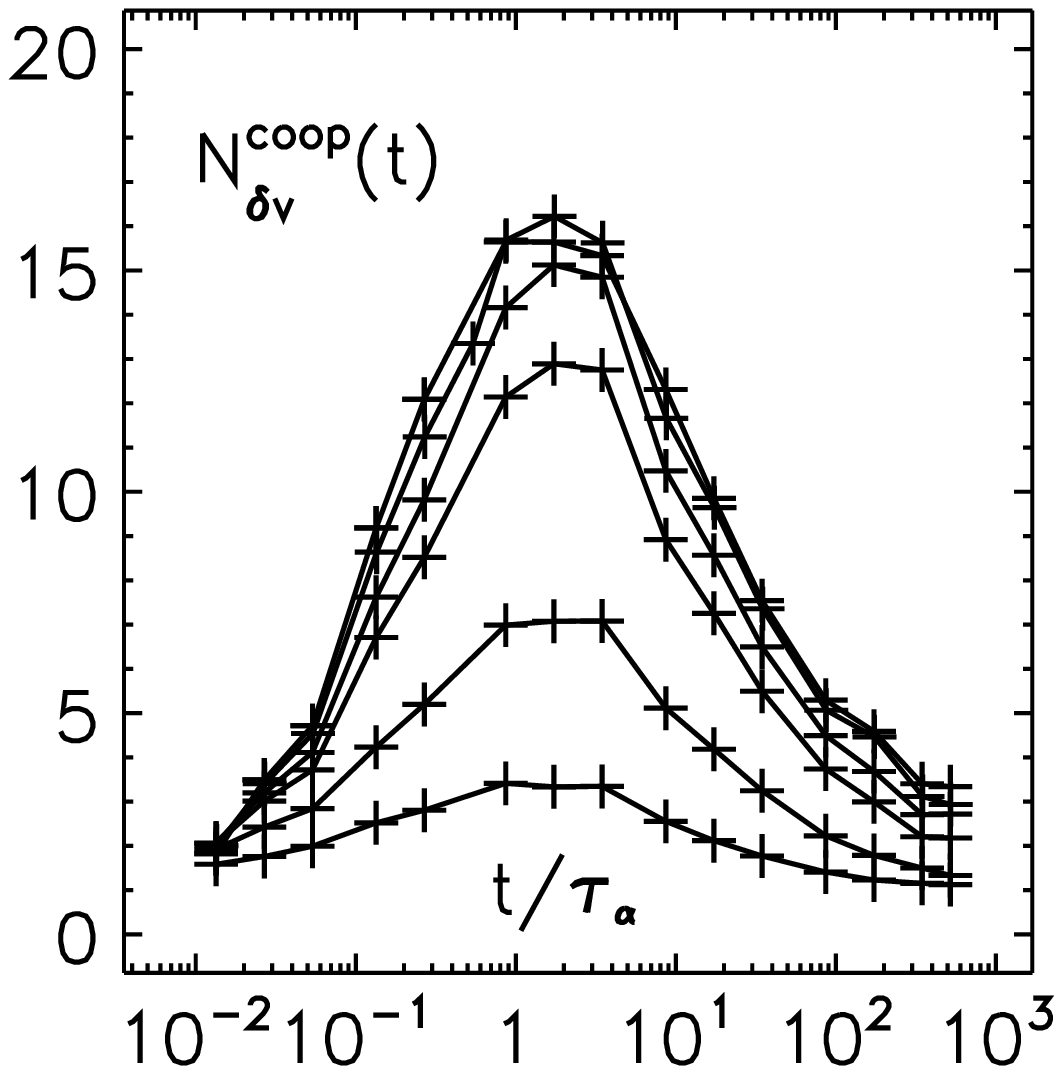}
\end{center} \vspm
}{
$\Ncodv(t)$ at density $\phithreed=0.56$, for different numbers n of
particles, that are summed over in $\sum_iX_i$.
From bottom to top: $n=20, 100, 500, 1000, 2000,$ and $4000$.
Time is normalized to \Mta.
}{ncoopN}

Naturally, a too small $n$ will
modify our results because we throw away some longer-ranged
correlations, which can be important.
An illustration for that is given in \FIG{ncoopN}, where $\Ncodv$
belonging to $X_i=\delta v_i\equiv v_i-\la v_i\ra$ is plotted
for different $n$ at density $\phithreed=0.56$.
We clearly see that it is necessary to take
as many as $n>1000$ particles because a major change can be found
when decreasing $n$ from there.
Interestingly, we obtained for a small system of $N=1066$, $\phithreed=0.56$
only a maximum $\Ncodv$ of nine when using $n=1066$ for its calculation
(data not shown).
This is half of the value of $\Ncodv$ at $N=16307$, $n=1000$,
thus proving large finite size effects in many-particle correlations
for the small system.
Being conscious of this problem is especially important if one
needs trustable numerical values for $\Ncoop$, e.g. for determining
the exponent of divergence when cooling towards the glass transition,
as is done in \cite{Novikov:1999}.
How is it possible that $\Ncodv<20$ in the case $\phithreed=0.56$
although the finite-size
effects of a too small subsystem $n$ can be sensed even up to $n=1000$?
The reason is that particles are only partially correlated, which will
become clear in the next section where we demonstrate the decay of
correlations with interparticle distance.
Additionally, regions of fast particles are extended, non-compact objects
(\FIG{MOVES}),
so that we have to sum over larger subsystems to include all their
mobility correlations.

On the other hand, however, taking $n$ as large as possible is not always
the best thing to do.
We can see the reason for this at the example $X=\mb{v}$ taking
$n=N$, i.e. all particles of the system.
Then, $\Var{\sum\mb{v}_i}=0$ because the simulation conserves
the center of mass, which means it sets $\sum\mb{v}_i$ to zero.
Fortunately, our systems are large enough, enabling us to choose
an optimum value of $n$ just between these two size effects.
For a more thorough discussion see \FIG{NCVinfty} and the
corresponding text below.

A word about error bars. The statistically uncertainty of simulation
results is a consequence of the limited length $T$ of the runs.
If we assume the equivalence of ensemble and time average, a quantity
$A$ can be determined up to accuracy $\var_T[A]\sim T^{-1}$ where
the constant of proportionality is essentially the
decay time of the autocorrelation function $\la A(t)A(0)\ra$.
We calculate $\var_T[A]$ by extrapolating its behavior for $T'<T$
to $T$, i.e. the average over the whole simulation run.
This is done for $A\equiv\Ncoop(t)$, only for few examples of $t$.
The resulting errors $\delta\Ncoop=\pm\left(\Var{\Ncoop}\right)^\frac12$
are given in the figure captions.

\figany{!ht}{
\centering\includegraphics[height=6.5cm]{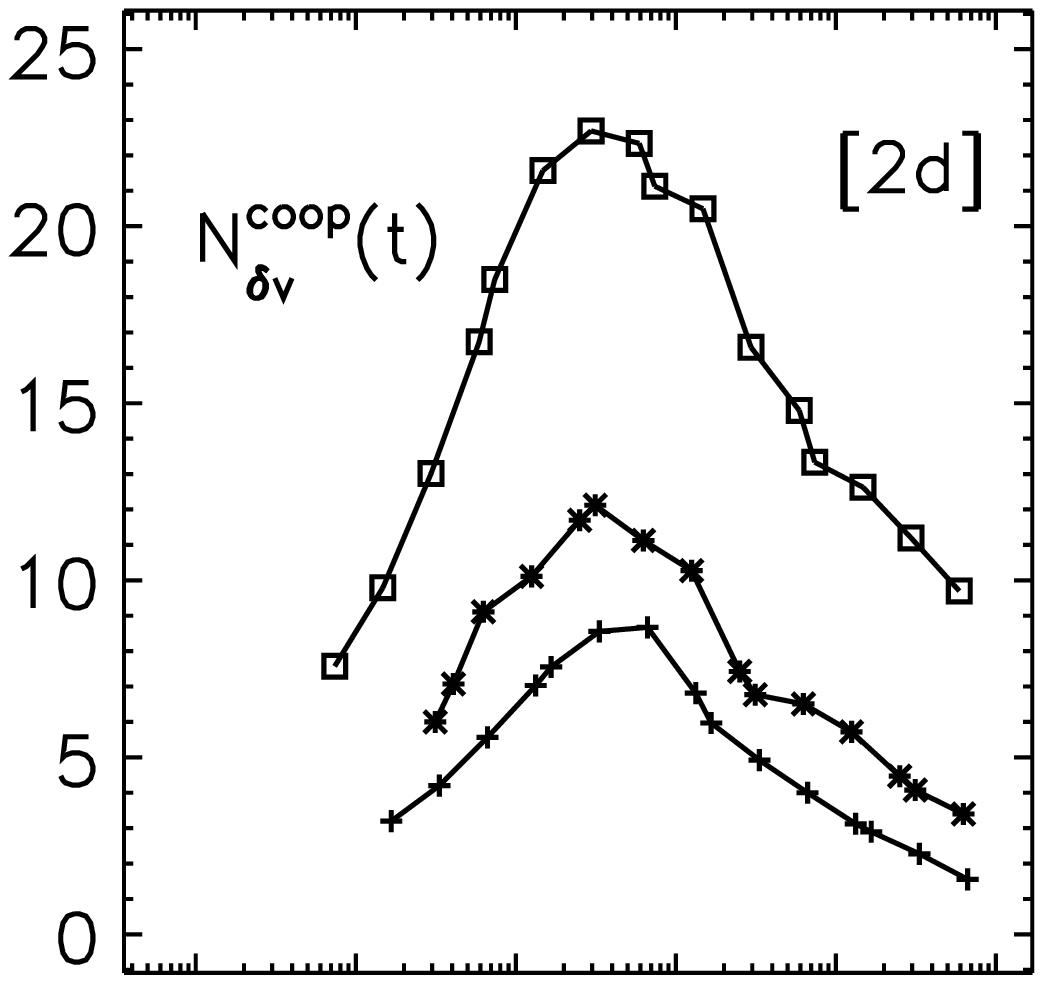}\\
\vspace{-1.5cm}
\centering\includegraphics[height=6.5cm]{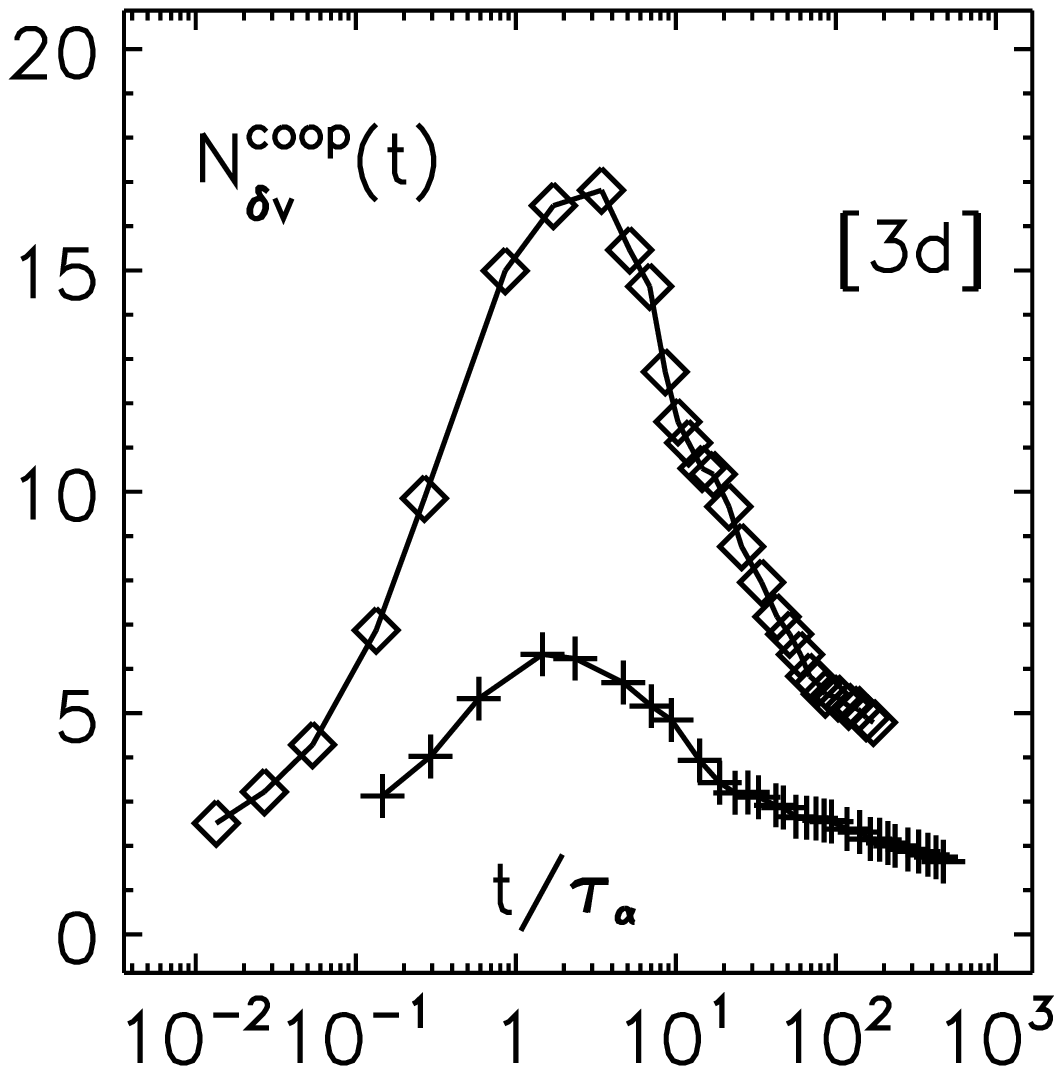}\\
\vspm
}{
[2d]:$\Ncodv(t)$ for $n=2000$ at $\phitwod=0.73, 0.75$ and $0.77$
[3d]:The same for the 3d case, $\phithreed=0.53$ and $0.56$.
The statistical error due to the finite simulation time is
$\delta\Ncodv(20\ta)=\pm0.5$ for $\phithreed=0.56$.
Again, time-normalization by \Mta.
}{ncoop2d3d}

\FIGU{ncoop2d3d} shows $\Ncoop=\Ncodv$ for some densities
$\phi$ in two and three dimensions.
As a function of time, $\Ncodv$ starts at short times with the value
of one because the individual brownian motions in the microscopic regime
are uncorrelated.
This is a trivial statement, so we do not have to demonstrate it
for every density.
For later times, $\Ncodv$ reaches a maximum which is strongly increasing
with density.
The following decay then takes some decades in time again.
But as can be seen, a limiting value is hard to observe within simulation times,
see below.

Now, the behavior of 2d and 3d systems seems to be quite similar,
although the maximum values of $\Ncodv$ are
larger in 2d.
A small difference is the shift of the 2d maxima towards longer times.
While in 3d they are found at approximately $2\ta$, we find them
in the 2d case at around 4$\ta$.
This shift can be observed in other dynamical quantities as well.
The simple reason lies in the different polydispersities $\sigPoly=0.1$ in three
and $\sigPoly=0.25$ in two dimensions.
In 2d, the small and - on average - fast particles
cause $F_2(\mb{k}_{\tiny\textnormal{max}},t)$ to decay more quickly in the beginning,
so we measure a systematically smaller $\ta$ than in the 3d case.
If we defined $F_2(\mb{k}_{\tiny\textnormal{max}},\ta)=0.01$
instead of requiring $F_2(\mb{k}_{\tiny\textnormal{max}},\ta)=\frac1e$,
this discrepancy would vanish.

\figany{!ht}{
\centering\includegraphics[height=6.5cm]{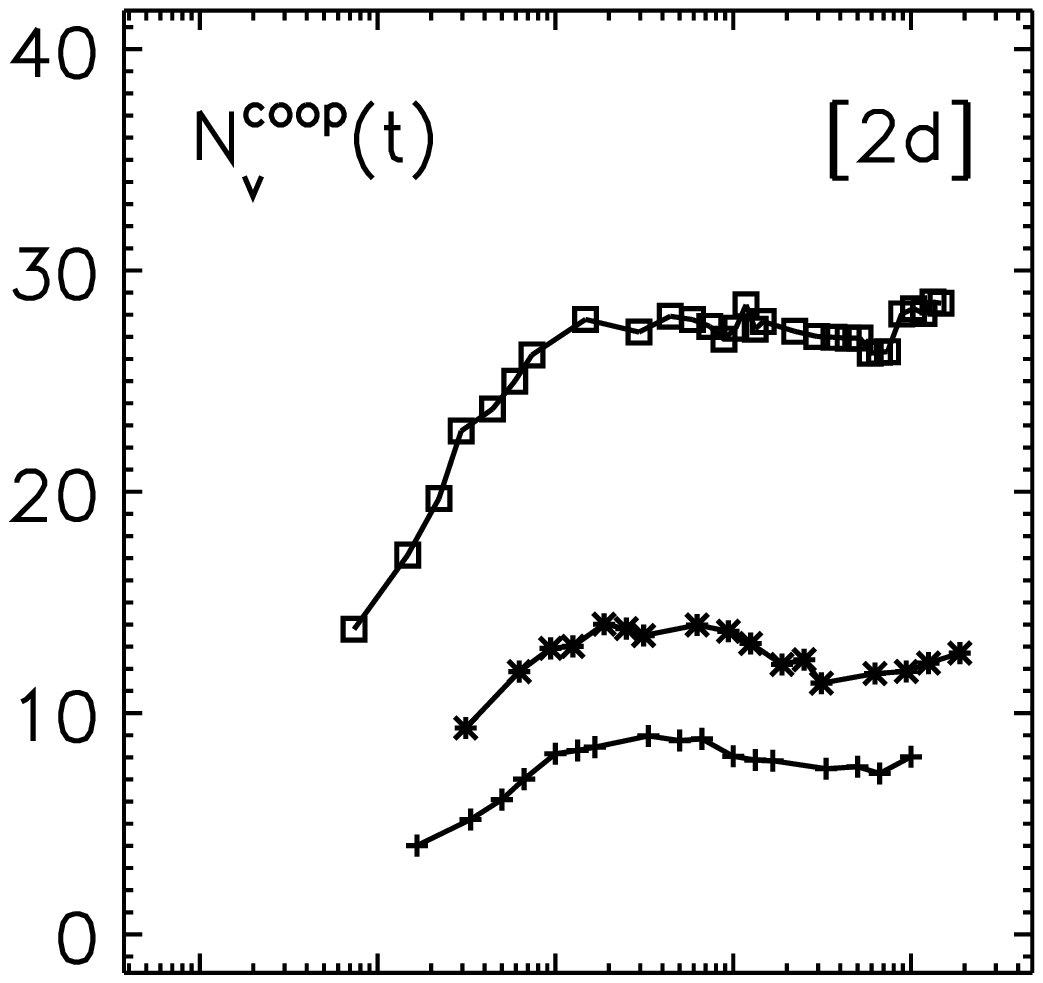}\\
\vspace{-1.5cm}
\centering\includegraphics[height=6.5cm]{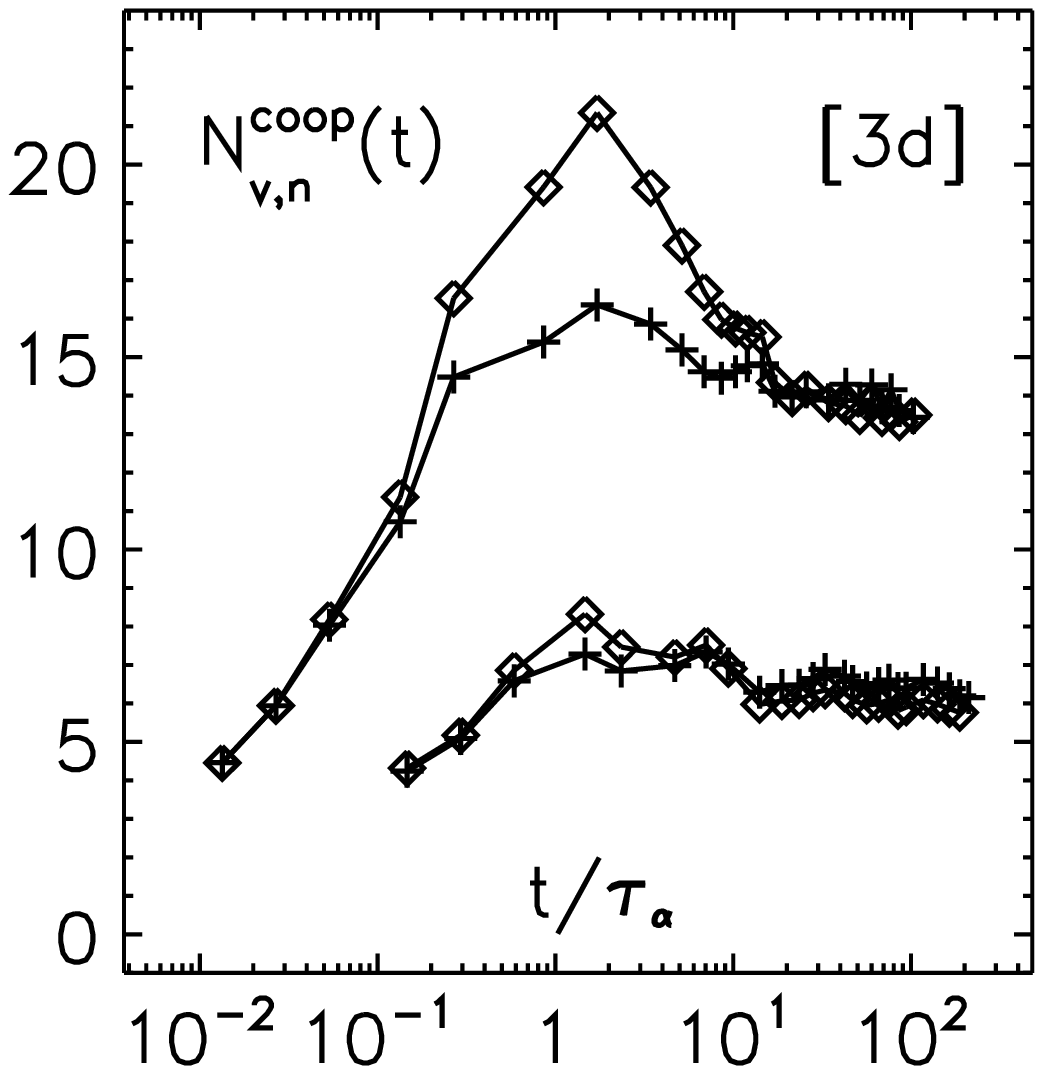}\\
\vspm
}{
[2d]: $\Ncov(t)$ for $n=2000$ at $\phitwod=0.73, 0.75$ and $0.77$.
[3d]: The same for the 3d case, $\phithreed=0.53$ and $0.56$,
$n=2000$ (+).
Additionally, we see $\Ncon(t)$ ($\Diamond$).
The statistical error is $\delta\Ncov(10\ta)=\pm0.5$ at $\phithreed=0.56$,
and $\delta\Ncov(50\ta)=\pm1.3$ at $\phitwod=0.77$,
for example.
Time is normalized by \Mta.
}{ncoopalt}

It is important to compare these results with $\Nco{,X}$ determined
by the dynamical quantities $X=\mb{v}$ or $X=\mb{n}$ as mentioned above.
For the interpretation of values of $\Ncoop$, this is essential,
because different sensible quantities $X$ should not produce totally
different values of \MNcoop.
\FIGU{ncoopalt} shows the case $X=\mb{v}$ for 2d and 3d, and
$X=\mb{n}$ for 3d.
Firstly, $\Ncov$ and $\Ncon$ clearly display their long-time limits,
which are equal and different from one.
Secondly, $\Ncon$ still develops a maximum around $t=\ta$, while $\Ncov$
is a monotonously increasing function.
Table \ref{TNCOOP} summarizes the main results about $\Ncoop$.
\begin{table}[!h]
\caption{}
\label{TNCOOP}
\begin{center}
\begin{tabular}{|c|c|c|c|}\hline
$\phitwod$ & $\ln\frac{D_0}D$ & $\Ncodvmax$ &
$\Ncovi$\\\hline\hline
0.73 & 2.76 & 9 & 8 {\tiny(10)}\\\hline
0.75 & 3.41 & 12 & 13 {\tiny(17)}\\\hline
0.77 & 4.51 & 23 & 28 {\tiny(35)}\\\hline
\end{tabular}
\end{center}
\begin{center}
\begin{tabular}{|c|c|c|c|c|}\hline
$\phithreed$ & $\ln\frac{D_0}D$ & $\Ncodvmax$ &
$\Ncovi$ & $\Nconmax$\\\hline\hline
0.53 & 2.92 & 6.5 & 6 {\tiny(7)} & 8\\\hline
0.56 & 4.53 & 17 & 14 {\tiny(16.5)} & 21\\\hline
\end{tabular}
\end{center}
\end{table}
We can compare the maximum values of \MNcoop for equivalent densities
(e.g. equivalent in the sense of equal $\frac D{D_0}$) and find a somewhat higher
cooperativity in 2d.
The absolute values of $\NcoX$ for the different $X$ agree to a reasonable extent,
so that we are indeed allowed to interpret them in the sense of a reduction factor
for the degrees of freedom.

Let us now turn to the limiting value $\Ncovi$.
Although we find a random diffusion for every particle on a time scale $t\gg\ta$
as expressed by the diffusion law $\sqrr\sim Dt$,
we should generally not expect $\NcoX$ to be one because
the inter-particle correlations from shorter times are still accounted for in this
quantity.
This can clearly be seen in the following way: We decompose the displacement $\vi$
for $t\gg\ta$ into $M$ pieces, each of them corresponding to a time step $\epsilon=\frac tM$,
i.e. $$\vi=\sum_{m=1}^M\Dv im,$$ for particle i.
For simplicity, let us say that inter-particle correlations are neglible for
different time intervals, i.e. $\la\Dv im\Dv j{m'}\ra=0$ if $m\not= m'$.
In this case, we obtain
$$\Ncov(t)=1+\frac{M\sum_{ij}\la\Dv i1\Dv j1\ra}{NM\sum_i\la(\Dv i1)^2\ra},$$
where we exploited time translational invariance, i.e. having an equilibrium liquid.
This quantity, however, does not depend on time $t=M\epsilon$ anymore,
if $\epsilon$ is fixed.
We thus
get an idea how it is possible that correlations persist for $t\to\infty$.

Finally, other choices of $X_i$ are possible, e.g.
more exotic quantities like
\EQU{X_i=w_i\equiv\left\{\begin{array}{r@{\quad:\quad}l}
		1 & \tm{slow} \\ -1 & \tm{fast} \\ 0 & \tm{otherwise}\end{array}\right.,}{}
where the exact definition of {\it fast} and {\it slow} is of no importance
as long as it is done in a sensible way.
Such an analysis has been presented in \cite{Novikov:1999} for a Lennard-Jones
fluid using a 'dynamic susceptibility' $\chi_\ind{SS}$ instead of $\Ncoop$.
Its definition is quite similar to $\Ncoop$, measuring fluctuations of
a many-particle, 'macroscopic' dynamic quantity $Q_\ind{SS}=\sum w_i$:
\EQU{\chi_\ind{SS}=\frac{\beta V}{N^2}\left[\la Q_\ind{SS}^2\ra-\la Q_\ind{SS}\ra^2\right].}{}
Unlike $\Ncoop$, a quantitative interpretation of the value of $\chi_\ind{SS}$
is not obvious.

\figany{!ht}{
\includegraphics[width=7cm]{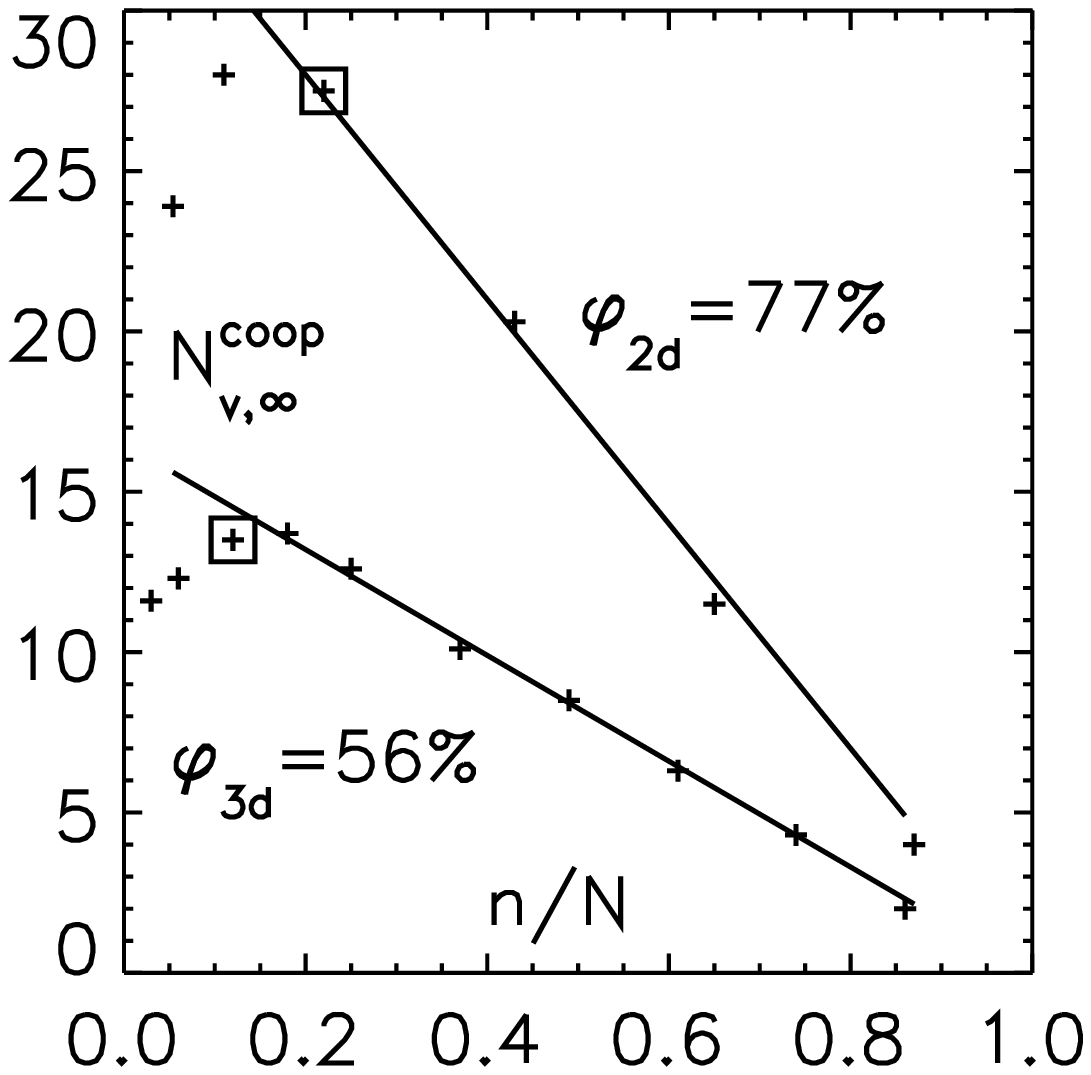}
\vspm
}{
The role of the subsystem size $n$ in the calculation of
$\Ncovi$ (+), for the systems
$\phitwod=77\%\ (N\!=\!9201)$ and $\phithreed=55\%\ (N\!=\!16307)$.
The solid lines are fits of the form
$\Ncovr(1-\frac nN)$, where the parameter
$\Ncovr=35$ and $16.5$, in the 2d and 3d case, respectively.
The $\Box$ marks the choice $n=2000$.
}{NCVinfty}
\renewcommand{\k}{\mb{U}}
\newcommand{\wj}{\mb{w}_j}
\newcommand{\wk}{\mb{w}_k}
\newcommand{\wl}{\mb{w}_l}
Let us return to the role of the size $n$ of the subsystems that are
used for the calculation of $\Ncoop$.
As we said above, the subsystem should be large enough to include most of
the long-ranged correlations of its particles, i.e. to reduce
surface effects. On the other hand, $n=N$ leads to $\Ncov=0$
because of the center-of-mass correction (cmc).
It is evident that our results will be influenced by the cmc
even if we use $n<N$, say $n=0.9N$.
Despite this fact, we need a cmc because the motion of the whole
simulation box leads to unphysical results for $\Ncoop$.
Interestingly, the center of mass performs a random walk with speed
$\eww{\vcm^2(t)}=\frac1N2dD_0t$,
independent of the packing fraction, which is the consequence
of $actio=reactio$ in a stochastic sense.
Now, the subtle point is that the cm motion consists of two contribitions,
firstly the random
displacement $\k$ of the simulation box as a whole which would vanish
if we embedded the simulation box in macroscopic system.
Secondly, the random rearrangements $\wi$ of particles in our system that
produce a contribution to the cm motion even if we forbid an
overall drift of the box.
Obviously, we should keep the second and
discard the first contribution because the latter is an artefact of
the limited system size.
A cmc, however, will remove both. In the following, we estimate
the resulting error in the calculation of $\Ncov$.
The uncorrected displacement of particle $i$ is
$\vi=\wi+\k$, so that the cm motion becomes
$\vcm=\frac1N\sum_{k=1}^N\wk+\k$, where the first term does
generally not vanish.
Calculating the numerator of $\Ncov$ in \REQU{GLNCOOP}, we obtain
\begin{eqnarray*}
& & \sum_{ij=1}^{n}\eww{\left(\vi-\vcm\right)\left(\vj-\vcm\right)} \\
&=&\sum_{ij=1}^{n}(\eww{\wi\wj}
	+\frac1{N^2}\sum_{kl=1}^N\eww{\wk\wl}\\
& &	-\frac2N\sum_{k=1}^N\eww{\wi\wk})\\
&=&\sum_{ij=1}^{n}\eww{\wi\wj}\left(1-\frac nN\right),\\
\end{eqnarray*}
where, in the final step, correlations between
the $\frac Nn$ different subsystems have been neglected.
Thus, \NEQU{\Ncov=\Ncovr\left(1-\frac nN\right)}{NCOVR}
is the result, which is too small by a factor of $(1-\frac nN)$.
As is demonstrated in \FIG{NCVinfty} for the value of $\Ncov(\!\infty\!)$,
this behavior can indeed be observed
in our simulations. The renormalized value of cooperativity, $\Ncovr$,
comes out as a fit parameter to the form of \REQU{NCOVR},
see table \ref{TNCOOP} in the brackets.
Additionally, \FIG{NCVinfty} tells us what choice of $n$ is advisable
because $\frac nN$ has to be in a region where the above fit works.
If this is the case, the subsystem must have only small
remaining correlations
with the other $N-n$ particles because increasing $n$
only reduces $\Ncov$ by the 'trivial' factor of $(1-\frac nN)$.
The choice $n=2000$ is marked by $\Box$ in \FIG{NCVinfty}.

In our case, at $\phitwod=0.77$, we observe a maximum reduction of the total degrees
of freedom by a factor of $\Ncovr=35$.
This, however, is yet only a moderately high density
($\phi_{\tm{\tiny{c,2d}}}\approx0.8$), so we should
expect large collective effects at the glass transition.

Finally we mention an interesting relation between
$\Ncovi$
and the Haven ratio, relating the ratio of the self-diffusion constant and the conductivity
in ionically conducting materials. Its zero-frequency limit $H(0)$ is given by
\begin{equation}
H(0) = \frac{\sum_{i} \int_0^\infty dt \langle \vi(0) \vi(t) \rangle }
           {\frac1N\sum_{i,j} \int_0^\infty dt \langle \vi(0) \vj(t) \rangle}.
\end{equation}

Since for large times $\langle \vi^2(t) \rangle = 2dDt = 2t \int_0^\infty dt \langle \vi(0) \vi(t) \rangle $
and correspondingly $\langle \vi(t) \vj(t) \rangle = 2t \int_0^\infty dt \langle \vi(0) \vj(t) \rangle$,
it is obvious that
\begin{equation}
\Ncovi = H(0)^{-1}.
\end{equation}
Hence, as a side product
we have obtained a quantitative interpretation of the inverse Haven ratio as the reduction of the effective degrees of
freedom.

\section{Spatial Correlations}

The snapshot of the dynamics in 2d (\FIG{MOVES}) demonstrates
that large spatial correlations are present in our systems.
In the following, we want to quantify them as a function of the timescale
of dynamics.

As in the treatment of $\NcoX$, a dynamical variable $X_i$
should be given for each particle, again with the restriction $\la X_i\ra=0$.
A spatial correlator can then be defined by
\NEQU{\la X(\mb{0})X(\mb{R})\ra\equiv
	\left\la\frac{1}{N}\sum_{ij}X_iX_j
		\delta\left(\mb{R}\!-\!(\mb{r}_i\!-\!\mb{r}_j)\right)
	\right\ra.
}{CORRDEF}
Again, $X_i$ denotes a dynamical quantity connected to the motion of
particle $i$ during the time interval $[t_0,t_0+t]$.
Because of symmetry reasons, the positions
$\mb{r}_i=\mb{r}_i(t_0\!+\!\frac t2)$ are used.
Averaging over the solid angle of $\mb{R}$, i.e.
\EQU{
\la X(0)X(R)\ra\equiv\frac1{4\pi R^2}\int{\rm d}
	\Omega\la X(\mb{0})X(\mb{R})\ra,
}{}
results in a loss of information because the direction of motion of
particle $i$, for example, breaks the isotropy.
\figany{!ht}{
\includegraphics[width=7cm]{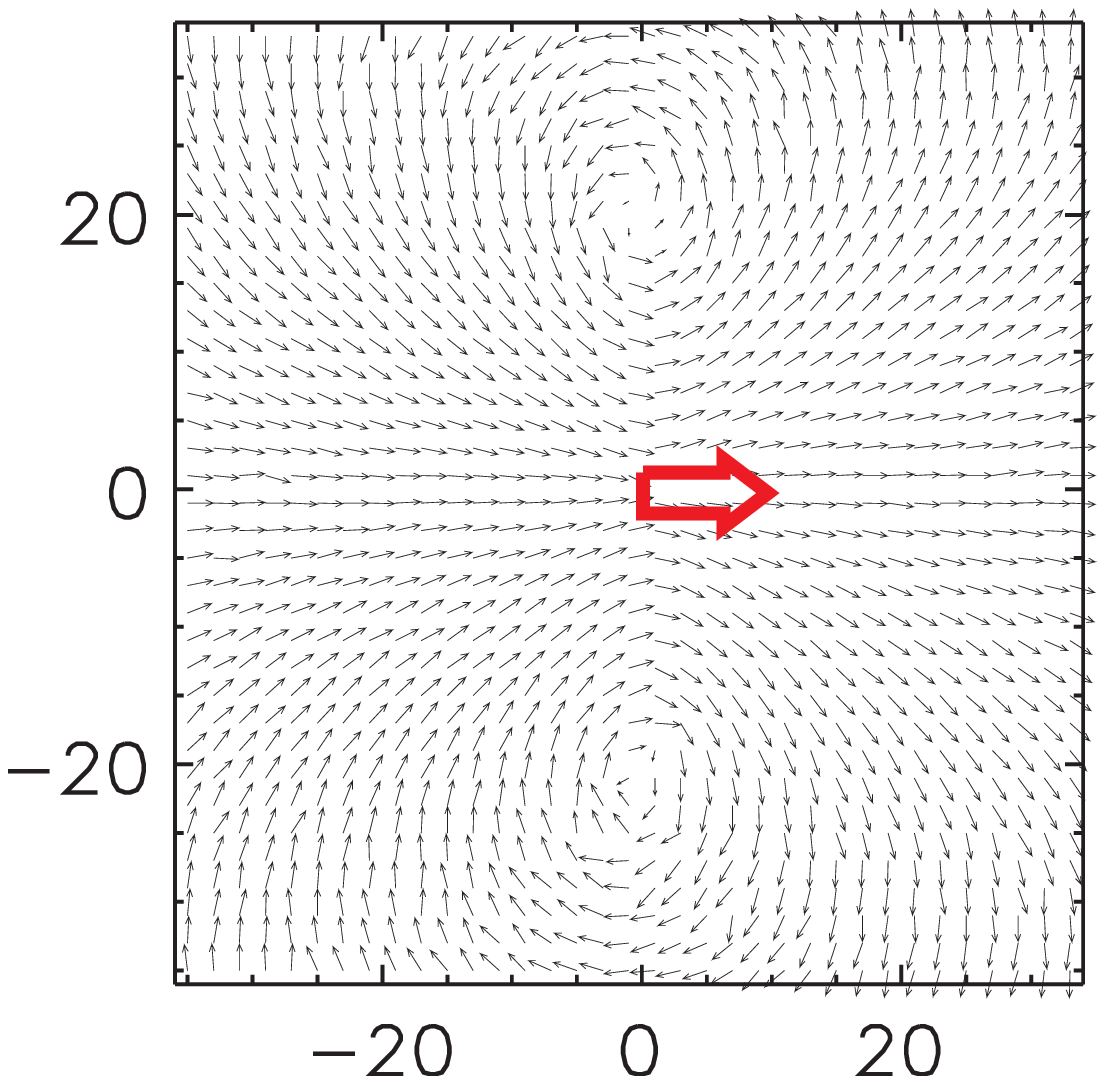}}{
Correlations of the direction of displacements, $X_i=\mb{n}_i$ at
$\phi=77\%, t=10\ta$. The large arrow in the middle shows the direction of
motion of the reference particle. For details, see text.
}{VCPIC}
\FIGU{VCPIC} explains this in a pictorial way:
particles "in front of" or "behind" the central one have a very
long-ranged directional correlation, while perpendicular to the
direction of motion, we observe a kind of back-flow behavior
which is well-known from \cite{Alder:1970}.
How has this plot been produced?
First, we calculate all $X_i=\mb{n}_i$, i.e. the directions of displacements
during a time interval $[t_0;t_0+t]$.
We then choose particle $i$ and turn the whole (2d) system so that $X_i'$
points along the positive x-axis. Now, the directions $X_j'$ are added
to the average at the positions $\mb{r}_j'(t_0+\frac t2)$.
As result, we obtain the field $\la X'(\mb{r}')\ra$, which for large
$||\mb{r}'||$ consists of very short vectors. Hence,
for reasons of visualization, we plot the {\it normalized} version of
$\la X'(\mb{r}')\ra$ in \FIG{VCPIC}.

Being aware of the complicated behavior in \FIG{VCPIC},
let us for the moment and for simplicity
ignore the angle-dependence, and treat
correlations only as a function of interparticle distance $R$.
We define the dimensionless quantity
\NEQU{S_X(R,t)\equiv\frac{\la X(0)X(R)\ra}{\la X^2(0)\ra},}{SOFR}
where again its dependence on time-scale $t$ should be kept in mind,
just as in the case of $\Ncoop$.
Possible choices are $X_i=\delta v_i$, $X_i=\mb{v}_i$
or $X_i=\mb{n}_i$, where $\mb{v}_i=\mb{r}_i(t)-\mb{r}_i(0)$,
$\delta v_i=v_i - \la v_i\ra$ and $\mb{n}_i=\hat{\mb{v}_i}$.
The functions $\Sdv$ and $\Sn$ count correlations of both slow and fast
particles because both sorts are weighted similarly.
To be more precise, the slow particles are not suppressed as in the case
of $\Svv$.

Calculating $S_X(R,t)$, we encounter a problem that is related to
the system size. If we want the system's center of mass to be constant,
we have to correct the particles' motions.
But this introduces an anti-correlation of two formerly uncorrelated
particles. As a consequence, $S_X(R,t)$ will approach a negative
value for large $R$, instead of zero.
Without center-of-mass correction, we would measure a positive number in this limit
because the center of mass performs a slow diffusive motion,
see the previous section.
For the present systems, these offsets were smaller than $0.002$,
which is a small fraction of the amplitude of correlation.
Hence, their subtraction from $S_X(R,t)$ left the function nearly unchanged.

\figany{!ht}{
\includegraphics[width=6.5cm]{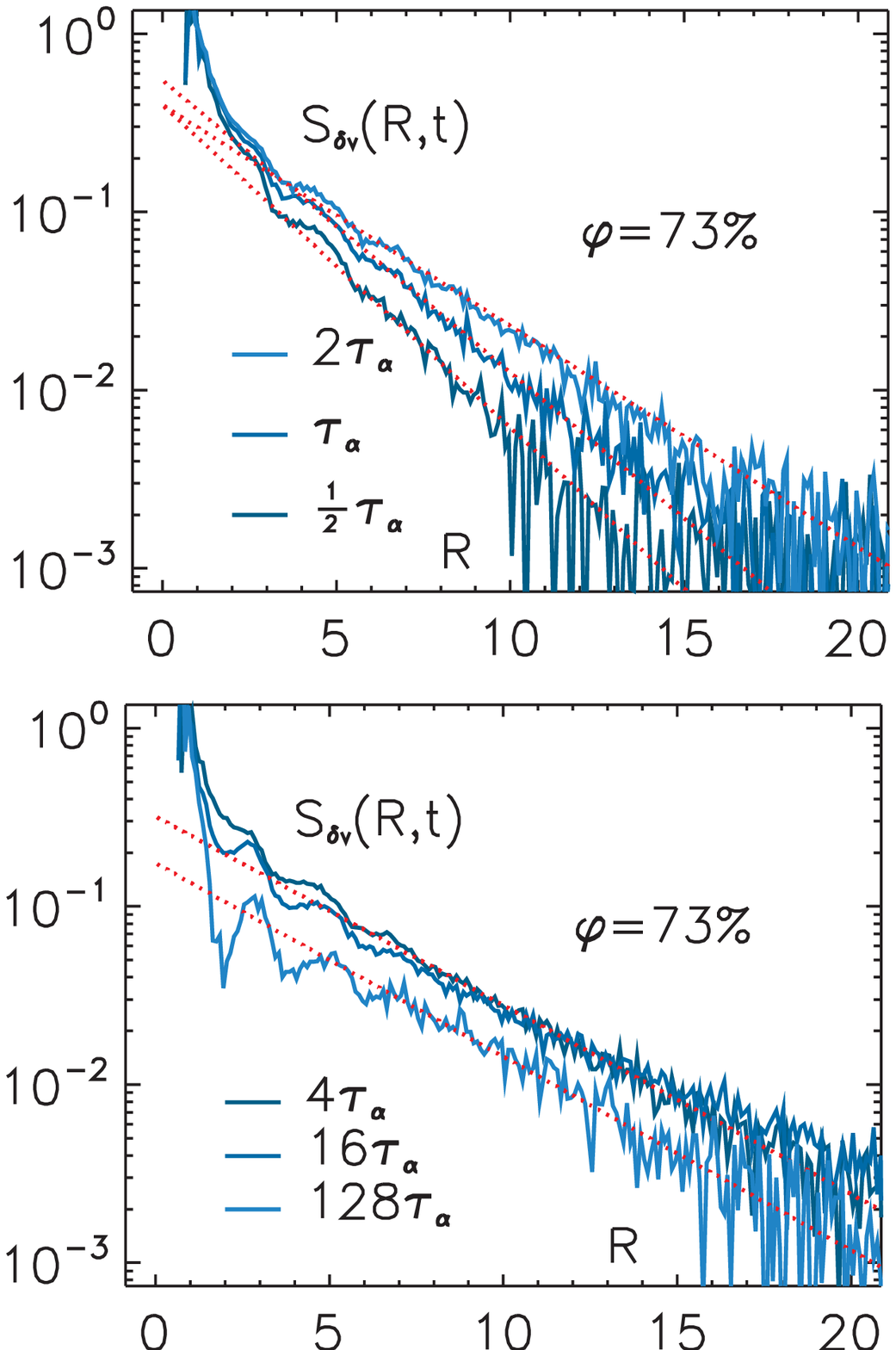}
}{Spatial correlation $S_{\delta v}(R,t)$ at $\phitwod=73\%$.
}{VCABS073}
\figany{!ht}{
\includegraphics[width=6.5cm]{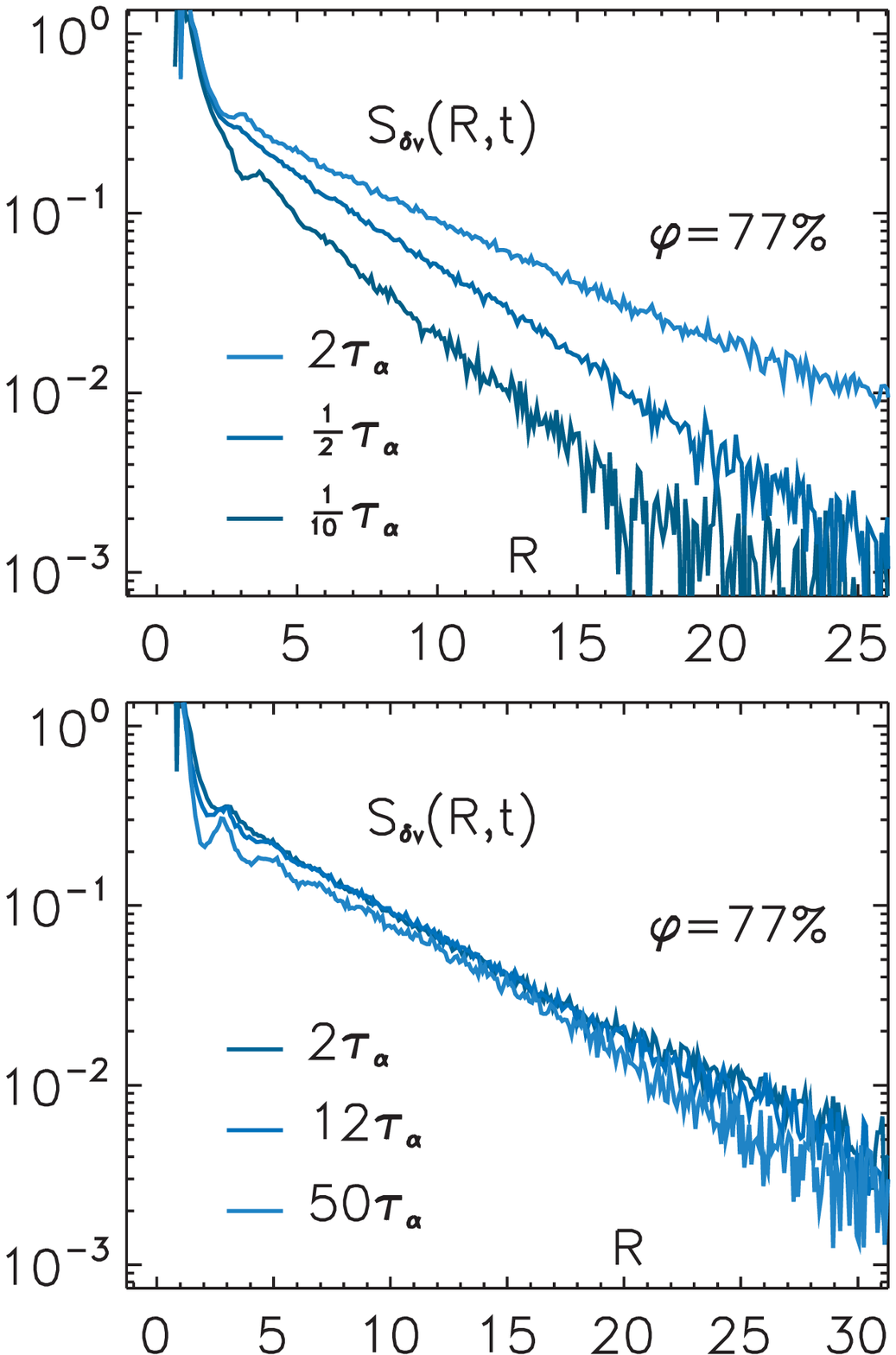}
}{Spatial correlation $S_{\delta v}(R,t)$ at $\phitwod=77\%$.
}{VCABS077}
\newcommand{\xiT}{\xi_X{(t)}}
\newcommand{\AT}{A(t)}
Figures \ref{VCABS073} and \ref{VCABS077} show $\Sdv$ for the 2d systems
$\phitwod=73\%$ and $\phitwod=0.77$.
Obviously, $\Sdv$ can be described
by an exponential
\NEQU{\Sdv\approx\AT\exp\left(-\frac R{\xidv(t)}\right)}{SEXPO}
to a good approximation, if $R>5$.
It is important to note that the amplitude $A(t)$ is not
necessarily equal to one as suggested by the definition of $S_X(R,t)$
for $R\to0$.
In other words, the extrapolation of complicated inter-particle
correlations to the one-particle quantity $S_X(R=0,t)$ would
be unphysical.

We find large deviations from the exponential at distances $R<5$.
This can be understood
qualitatively because certain information
about the local packing is available.
For instance, $R<1$ can only occur for two very small particles
(remember $\la R_i\ra\!=\!1$) which on average are much faster than the others.
Thus, $\la\delta v(0)\delta v(R)\ra$ will be quite large.
The oscillations for $R<6$ must have a similar reason, i.e. special local packings
that are favorable or not for the value of $\la\delta v(0)\delta v(R)\ra$.
We can imagine that for larger R, the possibilities
of packing become so many that they average $S_X(R)$ to a structureless
exponential. This is the case for the structure factor
$g(R)$, too.

\figany{!ht}{
\includegraphics[width=6.5cm]{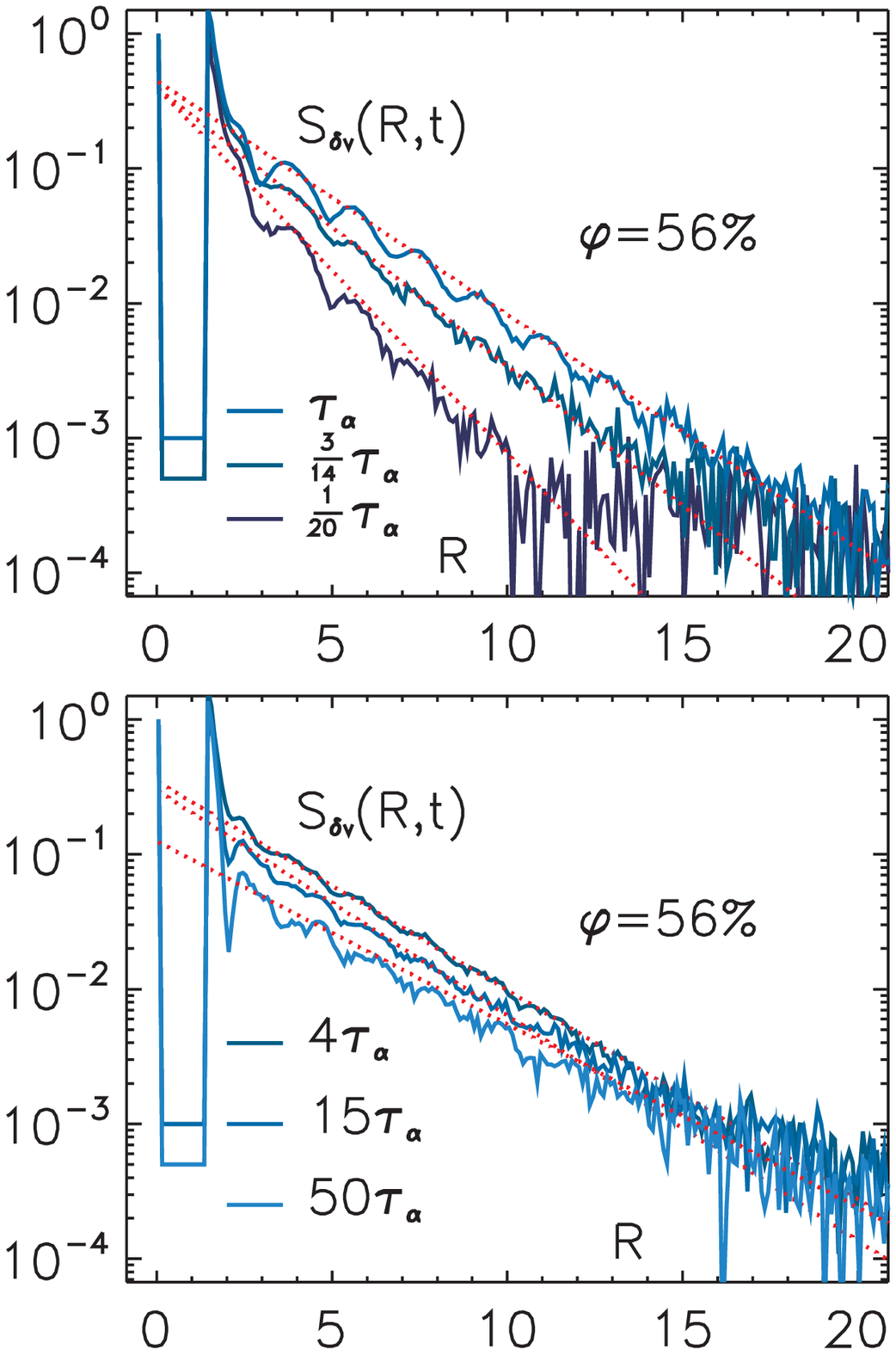}
}{Spatial correlation $S_{\delta v}(R,t)$ at $\phithreed=56\%$.
}{VCABS056}

\FIGU{VCABS056} shows $\Sdv$ for the three-dimensional case.
The situation is quite the same as in 2d, except for the long-range
oscillations, especially of $S_{\delta v}(R,t\!=\!\ta)$.
They, too, indicate the existence of structures that are favourable
for dynamical correlations.
For 2d systems of smaller polydispersity $\sigPoly=10\%$, which are
not shown here, we find the same oscillations.
In this case, they could be proved to result from local crystalline order,
which occurs, if - by coincedence - many particles of approximately
the same size come together. Although the system is in an overall
amorphous state, the small polydispersity makes local crystallites
more probable, thus creating regions of low mobility.
The oscillations for three dimensions, however, are not understood yet.

In any event, we can extract from $\Sdv$ the amplitude $A(t)$
and the correlation length $\xiT$ as a function of the dynamic
time scale $t$.
The simulation runs, by the way, have to be much longer than the maximum
time scale shown because the functions $\Sdv$ are quite demanding with respect to
statistics. For instance, distant particles, which are uncorrelated,
have to average $\la X(0)X(R)\ra$ to zero. The statistics $M$ enters
by a factor of $\frac{1}{\sqrt M}$, so an improvement of the result
has a high price.
Additionally, the {\it dynamic heterogeneities},
as visualized in \FIG{MOVES}, are very
long-lived \cite{HETEROLONGLIVED}, i.e. possess typical life-times
of tens to hundreds of $\ta$, dependent on $\phi$.
Thus, if we want to average over different dynamical situations, we need
data for some hundreds of $\ta$.
Table \ref{TSIMS} shows the lengths of our simulation runs
in units of \Mta for
the analyzed 2d and 3d packing fractions, respectively,
including the system size $N$.
\begin{table}[!h]
\caption{}
\label{TSIMS}
\begin{center}
\begin{tabular}{|c||c|c|c|c|}\hline
$\phitwod$ & 0.73 & 0.75 & 0.77 & 0.78\\\hline
\#$\ta$ & 2000 & 5000 & 5000 & 500\\\hline
$N$ & 8756 & 8960 & 9201 & 9320\\\hline
\end{tabular}
\end{center}
\begin{center}
\begin{tabular}{|c||c|c|c|c|}\hline
$\phithreed$ & 0.53 & 0.56\\\hline
\#$\ta$ & 1500 & 750\\\hline
$N$ & 15422 & 16307\\\hline
\end{tabular}
\end{center}
\end{table}

\figany{!ht}{
\includegraphics[width=6.5cm]{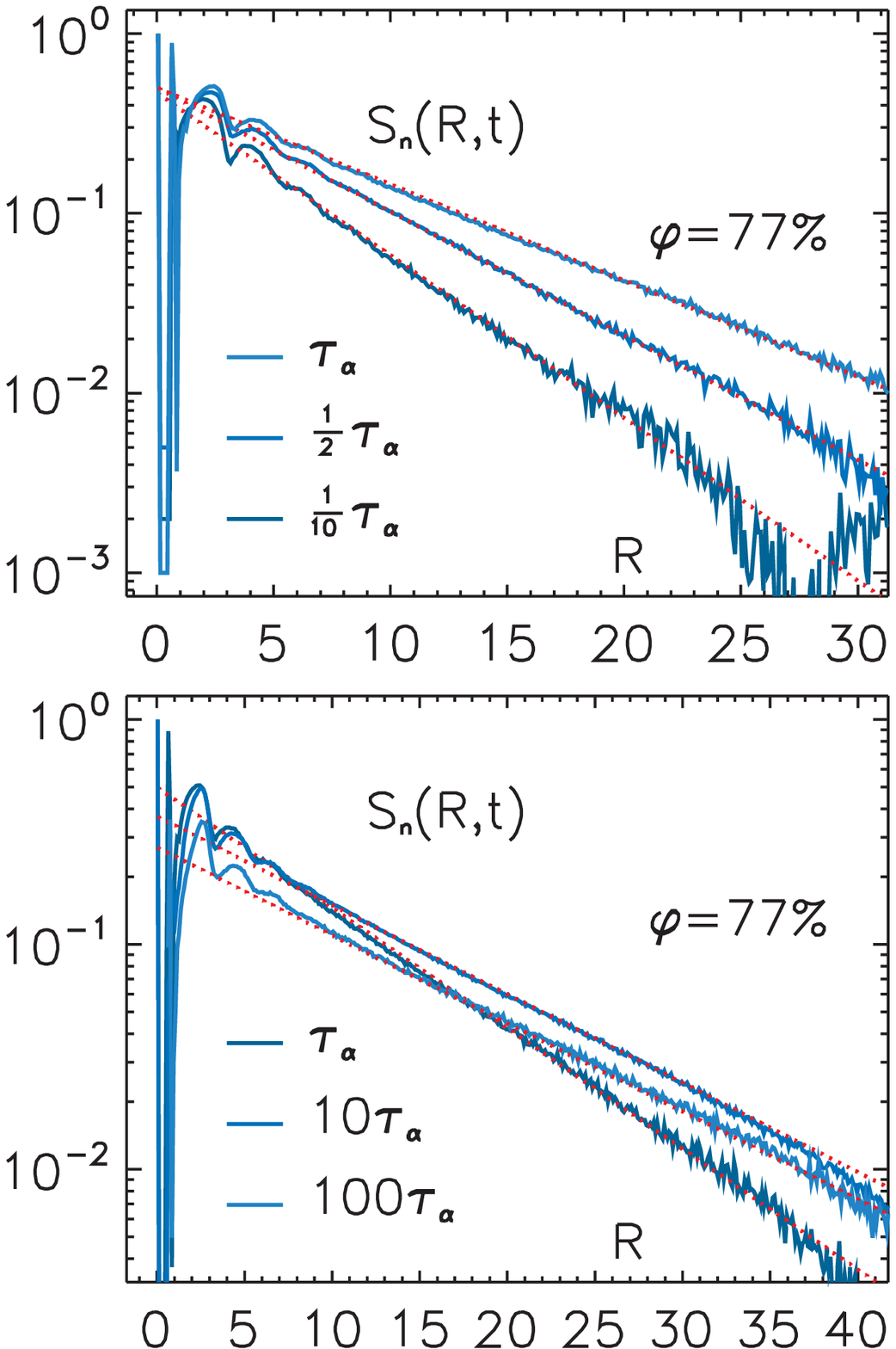}
}{Spatial correlation $\Sn$ at $\phitwod=77\%$.
}{VCDIR077}

\figany{!ht}{
\includegraphics[width=6.5cm]{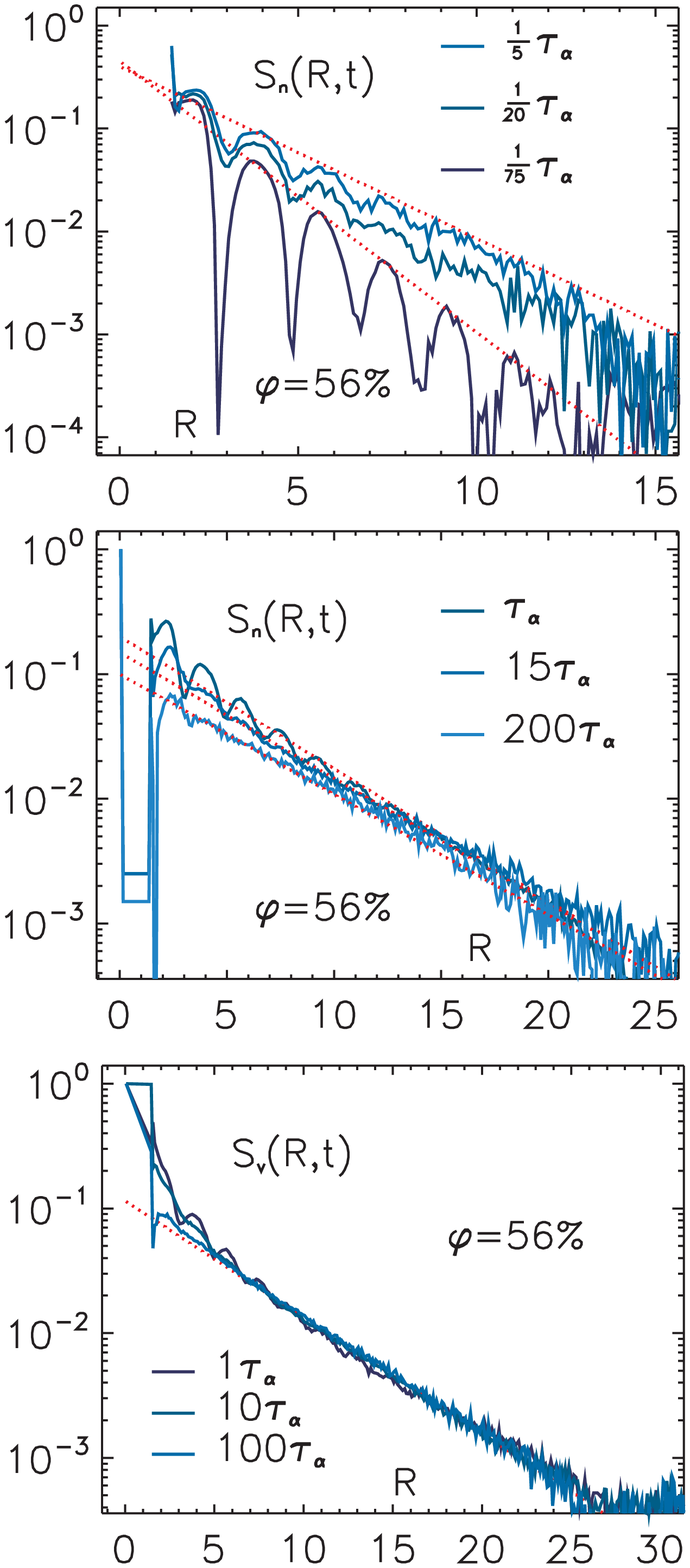}\vspace{.5cm}
}{Spatial correlation $S_{\mb{n}}(R;t)$ and $S_{\mb{v}}(R;t)$
at $\phithreed=56\%$.
}{VCDIR056}

Let us turn to another dynamical quantity $X_i=\mb{n}_i$, the direction of
displacement. As we see in \FIG{VCDIR077}, $\Sn$ is quite
similar to $\Sdv$,
i.e. we find an overall exponential decay
of correlations with distance $R$.
Its characteristic length $\xi_{\mb{n}}(t)$, however, is much larger
than the previous $\xi_\ind{$\delta v$}(t)$ (see below).
Again, the situation is quite the same in three dimensions (\FIG{VCDIR056}), i.e.
$\xi_\ind{$\mb{n}$}$ exeeds $\xi_\ind{$\delta v$}$ for $\phithreed=0.56$.
A remarkable difference to the 2d case are the extreme oscillations of $\Sn$ for
$t=\frac1{75}\ta$. This is not understood yet, but could be due to the lower
polydispersity in the three-dimensional system causing locally less amorphous
packings.
The vectorial correlation $\Svv$ is also shown in \FIG{VCDIR056}, where
it seems that the amplitude $A_{\mb{v}}(t)$ is a constant for all time scales
$t>\ta$.

\FIGU{LENGTHS} summarizes the data for $\xiT$, $X_i=\delta v_i$.
First of all, we
notice an increase of correlation lengths with density.
For a 3d Lennard-Jones system \cite{Poole:1998}
and polymers \cite{Bennemann:1999}, this has already
been demonstrated for the special choice of time scale $t\approx\ta$.
But for longer times, even larger correlation lengths can be observed
as shown in the figure.
Interestingly, $\xiT$ is a monotonously increasing function,
with a limiting value $\xiXi\equiv\lim_{t\to\infty}\xiT$.
For comparison, \FIG{LENGTHS} shows the length scales for the
directional correlation $\xin(t)$ at $\phitwod=0.77$ and
$\phithreed=0.56$.

Table \ref{TLENGTHS} summarizes the data for $\xiX(\infty)$, where
the error due to fitting the exponential is less than ten percent.
(The quantity $\Rcurl$ will be explained below.)
The statistical error due to finite time averages
is small enough to be included therein.
\begin{table}[!h]
\caption{}
\label{TLENGTHS}
\begin{center}
\begin{tabular}{|c||c|c|}\hline
$\phithreed$ & 0.53 & 0.56\\\hline
$\xidv(\infty)$ & 2.2 & 2.8\\\hline
$\xin(\infty)$ & 2.3 & 4.5\\\hline
$\xiv(\infty)$ & - & 4.6\\\hline
\end{tabular}
\end{center}
\begin{center}
\begin{tabular}{|c||c|c|c|c|}\hline
$\phitwod$ & 0.73 & 0.75 & 0.77 & 0.78\\\hline
$\xidv(\infty)$ & 4.1 & 5.2 & 6.0 & 7.2\\\hline
$\xin(\infty)$ & 3.7 & 5.4 & 11 & 16\\\hline
$\xiv(\infty)$ & 4 & - & 10.5 & 16\\\hline
$\Rcurli$ & 8 & 10 & 21 & 30\\\hline
\end{tabular}
\end{center}
\end{table}
As we see, the length scale $\xidvi$ takes a snuggish
rise, growing from $4.1$ to $7.15$ between $\phitwod=0.73$ and $0.77$.
In contrast, $\xini$ and $\xivi$, starting from about the same initial
value, end up at a value of more than twice $\xidvi$.
The underlying physics of this very different behavior of mobility
and directional correlations is unclear.
As in the case of $\Ncoop$, the vectorial quantities $\mb{n}_i$ and $\mb{v}_i$
show a very similar behavior of their inter-particle correlators.

So we can say, that the overall cooperativity is determined by both
its length scale $\xiT$ and its strength, or amplitude, $A(t)$.
Eqns. (\ref{GLNCOOP}) and (\ref{SOFR}) show that
\begin{eqnarray*}
\Ncoop	&\equiv&1+\frac{\sum_{i\ne j}\la X_iX_j\ra}{\sum\la X_i^2\ra}\\
	&=&\frac1{\sum\la X_i^2\ra}\sum_{ij}\la X_iX_j\ra\\
	&=&\frac1{\la X^2(0)\ra}\int_0^{\infty}{\rm d}Rp(R)\la X(0)X(R)\ra\\
	&=&\int_0^{\infty}{\rm d}Rp(R)S_X(R),\\
\end{eqnarray*}
where $p(R)$ is the average number of particles found at distance $R$ from
a particle at the origin.
In a way, this is a trivial result because the sum over all spatial
correlations {should} be the overall cooperativity.
When the approximation $S_X(R)\approx Ae^{-R/\xi}$ is valid and at
homogeneous density, we find $\Ncoop(t)\sim\xiT^3\AT$.
The deviations from the exponential for small $R$ can modify this argumentation,
but are unlikely to totally change the picture.

\figany{!ht}{
\includegraphics[width=5.5cm]{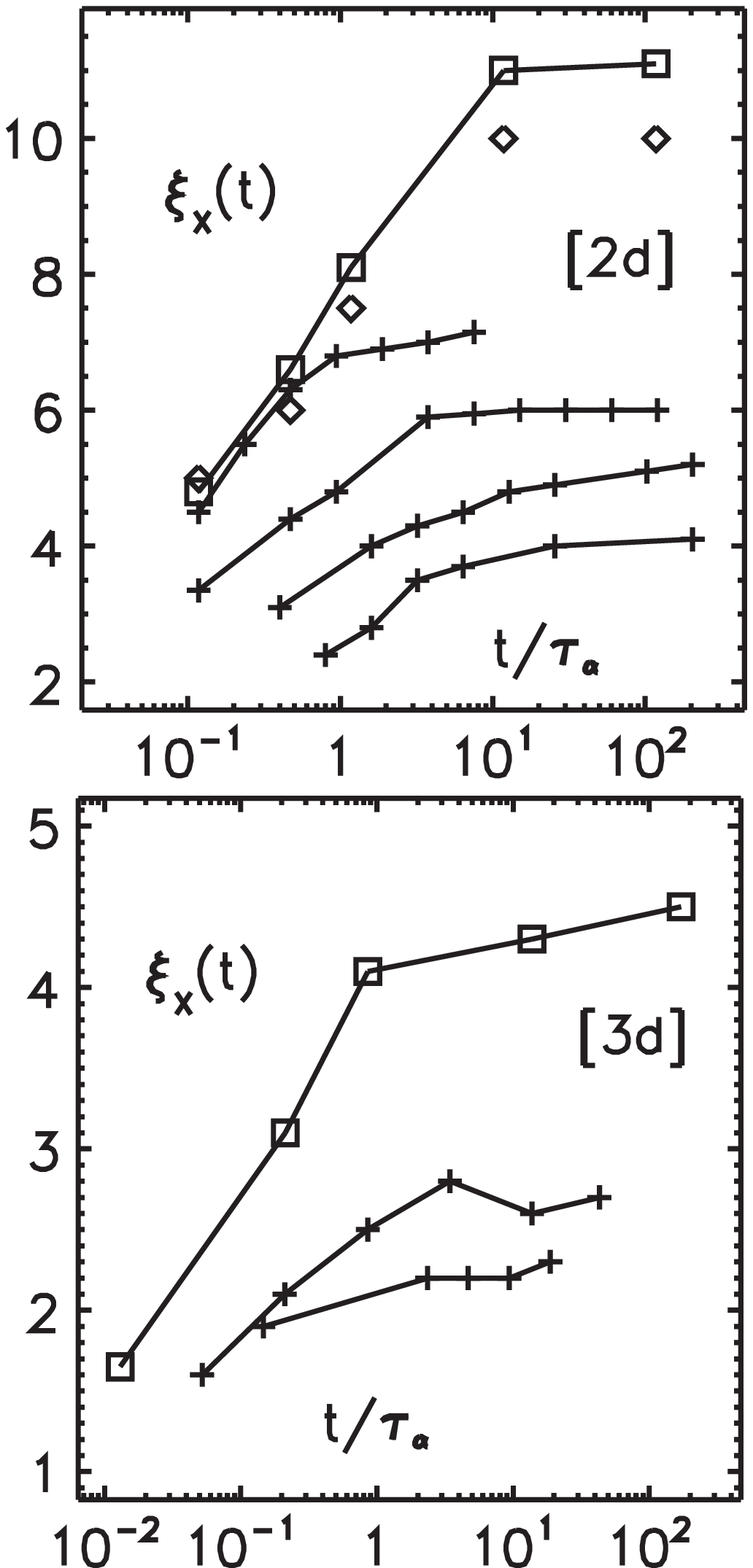}
}{[2d]: The dynamical length scales $\xidv(t)$
at packing fractions $\phitwod=73\%$, $75\%$, $77\%$ and $78\%$,
from botton to top (+).
For comparison: $\xin(t)$ ($\Box$)
and $\frac12\Rcurl(t)$ ($\Diamond$) at $\phitwod=77\%$.
[3d]: Again $\xidv(t)$ for the 3d densities $\phithreed=53\%$ and $56\%$
(+) and $\xin(t)$ at $\phithreed=56\%$ ($\Box$).
Errors due to fitting are estimated to be less than ten percent.
}{LENGTHS}

Consequently, the results for $\Ncodv(t)$ and $\xidv$ are only compatible,
if the strength of correlation $A_{\delta v}(t)$ will tend to zero for long
times.
We can observe the decrease of $A_{\delta v}(t)$ clearly
in \FIG{VCABS073} ($\phi=0.73$),
but it is harder to see at higher densities
because of the limited time window.
For $X_i=\vi$, in contrast, we need a limiting value of $A_{\mb{v}}$
greater than zero, if $\Svv$ is to be compatible
with $\Ncov(t)$ for $t\!\to\!\infty$.
\FIGU{VCDIR056} proves this to be the case because $A_{\mb{v}}(t)$ is
constant for $t>\ta$.

Let us finally return to the detailed picture of \FIG{VCPIC}.
It suggests, that spatial correlations {\it along}
the direction of motion will be very different from
them {\it perpendicular} to it.
We can test this by restricting the summation in \REQU{CORRDEF}
to certain angles $\psi$ between the motion of particle $i$ and the
connection vector $\mb{r}_j\!-\!\mb{r}_i$.
For example, the condition $\psi\in[\frac\pi3,\frac{2\pi}3]$ chooses
only particles $j$ that are {\it collateral} to particle $i$ with respect
to its motion $\mb{v}_i$.
To select {\it in direction} of motion, we demand
$\psi\in[0,\frac{\pi}{20}]\cup[\frac{19}{20}\pi,\pi]$, for example.
The restricted sum is then carried out to obtain $\Sperp$ or $\Spara$,
see \FIG{VCDIRPARSENK}. Because of the backflow, we expect
$\Spara$ to become negative for large $R>20$.
This can be observed in the figure.
In contrast, $\Spara$ approaches zero fo $R\to\infty$.
We notice the difference of length scales $\xiperp$ and $\xipara$,
which is summarized in Table \ref{TPARSENK}.
\begin{table}[!h]
\caption{}
\label{TPARSENK}
\begin{center}
\begin{tabular}{|c||c|c|}\hline
$\phi$ & 0.56 & 0.77 \\\hline
$\xiperp(\!\infty\!)$ & 2.7 & 7.5\\\hline
$\xipara(\!\infty\!)$ & 4.4 & 15\\\hline
\end{tabular}
\end{center}
\end{table}
The estimated error of these length scales is less than $20$ percent,
but the exact values are not of interest here.
Instead, the notion of a very different behavior of particles
parallel and perpendicular to the motion of the central one
is justified for three as well as for two-dimensional systems.

\figany{!ht}{
\includegraphics[width=6.5cm]{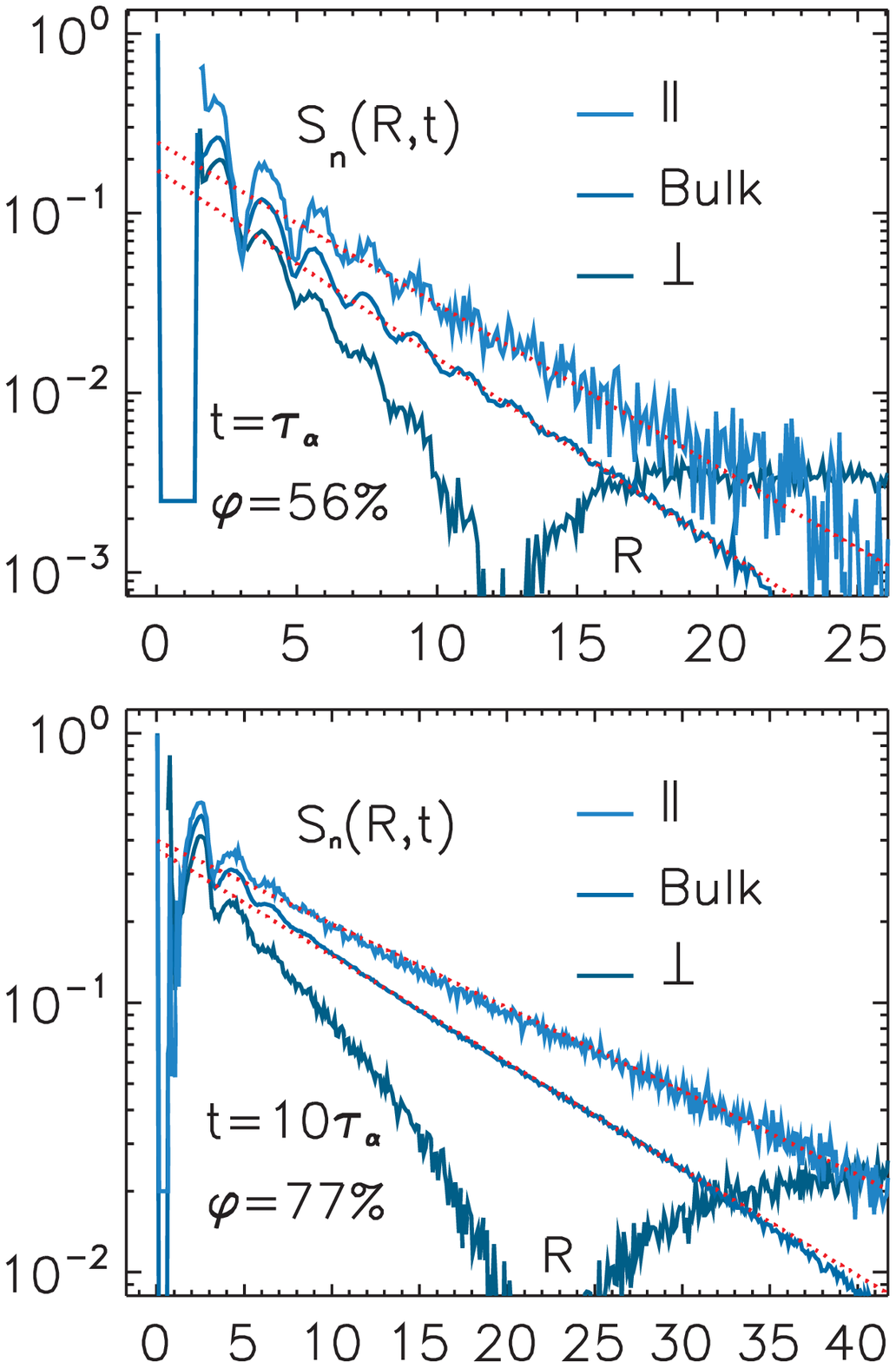}
}{Spatial correlation $\Sn$ at $\phitwod=77\%$, $t\!=\!10\ta$
and $\phithreed=56\%$, $t\!=\!\ta$,
calculated dependent on the angle $\psi=(\mb{v}_i, \mb{r}_j\!-\!\mb{r}_j)$.
Parallel ($||$) means
$\psi\in[0,\frac{\pi}{20}]\cup[\frac{19}{20}\pi,\pi]$,
perpendicular ($\perp$) means $\psi\in[\frac{\pi}3,\frac{2\pi}3]$, and
bulk stands for $\psi\in[0,\pi]$.
In the perpendicular case, $\Sn$ becomes zero at $R\approx\Rcurl$ and is
negative for larger distances, so that it is
necessary to plot its absolute value.
}{VCDIRPARSENK}

The time dependence $\xiT$,
if at first sight surprising,
can be understood quite pictorially.
Without any further information, the probability for
the motion of some tagged particle $i$ is equal in every direction.
However, if we know that, in the meanwhile, one or more of its
next neighbors perform some specified
displacements, this will influence the probability
of movement of the tagged particle.
More distant neighbors will do this as well, but the information
about their motions has to be 'submitted' to particle $i$
via nearer neighbors.
It is not hard to imagine that the information spread can only
take place with a finite velocity.
Thus, short-time motions will be accompanied by a reaction
of few neighbors, while long-lasting displacements will involve
many of them.
The monotonously growing length scale of dynamic correlations
is the natural consequence.

In the limit $t\!\to\!\infty$, we can argue as in the case of $\Ncoop$:
For long waiting times, the displacements
$\ui=\mb{r}_i(\frac t2)-\mb{r}_i(0)$ and
$\wi=\mb{r}_i(t)-\mb{r}_i(\frac t2)$
become independent to a good approximation.
Thus, correlations on time scale $t$ can be expressed through $\ui$
and $\mb{w}_i$, using $\vi=\ui+\wi$:
\EQU{\eww{\mb{v}(0)\mb{v}(R)}\approx\eww{\mb u(0)\mb u(R)}'+\eww{\mb w(0)\mb w(R)}''.
}{SPACECORRTTH}
This results in
\EQU{\Svv\approx\frac12 S^{'}_{\mb v}\left(R,\frac t2\right)
	+\frac12 S^{''}_{\mb v}\left(R,\frac t2\right).
}{}
The primed and double primed versions of $\Svv$, respectively,
denote measuring the distances $R$ at the end or at the beginning
of the time interval $[0,t]$.
On the other hand, $\Svv$ is defined by using the interparticle distance $R$
in the middle of this interval.
However, the definitions $\Svv$, $S^{'}_{\mb v}(R,t)$ and $S^{''}_{\mb v}(R,t)$
produced the same results in our simulations, which is not shown here.
Thus, in the long-time limit, the spatial correlations of the vectorial displacements
$X_i=\vi$ become time-independent, i.e.
\EQU{\Svv\approx S_{\mb v}\left(R,\frac t2\right).}{}

The phenomenon of a growing dynamical length scale
is exibited by much simpler systems, like a
one-dimensional (closed) chain of $N$ diffusive particles which are
connected by harmonic springs, as described by the Langevin equation
$$\dot{x}_n=-k(2x_n-x_{n-1}-x_{n+1})+\eta_n,$$
where the $\eta_n$ are independent white noises.
Let us assume that $N$ is a large number, say $N>1000$.
The analytic solution of this many-particle problem is possible
with the help of discrete Fourier transform. This enables us to calculate
the displacement-displacement correlation, but this is not shown here.
In this simple model, the length of correlation increases with time, too. 
In contrast to our simulations, $\xiv(t)$ grows until it has reached the system size.
Only from this time on
center-of-mass diffusion prevails, for which correlations among different particles
are no longer relevant.
Stated differently, apart from finite size effects the chain model possesses $\xivi=\infty$.
In turn, we see the reason for a finite value of $\xiXi$ in our HS systems:

{\it"Particles simply can go out of each other's way".}

In other words, a particle that travels a long distance does not have to
pull the whole system with it because rearrangements are possible.
On average, this results in the back-flow behavior of \FIG{VCPIC}.
We are therefore tempted to relate the length scales $\xiXi$ to an inherent length
of the back-flow pattern for long times.
From \FIG{VCPIC}, we see that the distance $\Rcurl(t)$
from the vortices to the central particle
is the only sensible choice.
Table \ref{TLENGTHS} shows the limiting values $\Rcurli$ for the 2d systems
under investigation.
Interestingly, $\Rcurli$ is twice the correlation length $\xivi$ or $\xini$.
For $\xidvi$, no such relation seems to exist.

\section{Discussion}

We presented detailed information about displacement correlations,
which turned out to be of the same nature for two-dimensional
discs and three-dimensional hard spheres.

Using the quantities $X_i=\delta v_i$, $\mb{n}_i$ and $\mb{v}_i$ as input
for $\NcoX(t)$ and $\SX$, we were able to measure the total reduction
of degrees of freedom and the spatial extent of correlations, respectively.
The data of $\NcoX$ is found to agree for these three choices of $X_i$,
supporting the notion of a reduced dimensionality of motion in
high-dimensional configuration space.
The length scale of correlations, however, turns out to increase much faster
with density for $X_i=\mb{n}_i$ and $\mb{v}_i$ than for $X_i=\delta v_i$.
An explanation for this is lacking at the moment.
Finally, we demonstrated, that for 2d as well as for 3d systems, an angle-resolved
calculation of correlations is appropriate, yielding much larger length scales
{\it in} the direction of motion than {\it perpendicular} to it.

The important question arises how the dynamical length scales $\xiXi$
are connected to static correlations.
Naive attempts, using, e.g. spatial density correlations, have not
revealed any significant connection.
What makes things more complicated is $\xiT$'s dependence on
the definition of $X_i$.
On the other hand, $\NcoX(t)$ and $S_X(R;t)$ are strongly
averaged quantities,
obtained by including many different dynamical situations.
Thus, we should not hope
to get too specific information from them.
A deeper understanding of cooperative effects will only become
possible by a more detailed, less averaged
treatment.
In any way, the relation of structure to dynamics
is the central problem to be solved.
The present work may help to formulate the relevant
questions somewhat clearer.

We gratefully acknowledge helpful discussions with K. Binder, S.
B\"uchner, M. Fuchs, J. Qian, B. Roling, and H.W. Spiess.
This work was supported by the DFG (SFB 262) and the Fonds der
Chemischen Industrie.

\bibliographystyle{unsrt}
\bibliography{Refs}

\end{document}